\documentclass[letter, superscriptaddress, prd, preprint, nobalancelastpage, nofootinbib]{revtex4}

\usepackage{verbatim, amsmath, amssymb, graphicx, color}

\allowdisplaybreaks

\begin{document}

\title{Nonperturbative analysis of the evolution of cosmological perturbations through a nonsingular bounce}

\date{}

\author{BingKan Xue}
\affiliation{Department of Physics, Princeton University, Princeton, New Jersey 08544, USA}

\author{David Garfinkle}
\affiliation{Department of Physics, Oakland University, Rochester, Michigan 48309, USA}
\affiliation{Michigan Center for Theoretical Physics, Randall Laboratory of Physics, University of Michigan, Ann Arbor, Michigan 48109, USA}

\author{Frans Pretorius}
\affiliation{Department of Physics, Princeton University, Princeton, New Jersey 08544, USA}

\author{Paul J. Steinhardt}
\affiliation{Department of Physics, Princeton University, Princeton, New Jersey 08544, USA}
\affiliation{Princeton Center for Theoretical Physics, Princeton University, Princeton, New Jersey 08544, USA}

\begin{abstract}
In bouncing cosmology, the primordial fluctuations are generated in a cosmic contraction phase before the bounce into the current expansion phase. For a nonsingular bounce, curvature and anisotropy grow rapidly during the bouncing phase, raising questions about the reliability of perturbative analysis. In this paper, we study the evolution of adiabatic perturbations in a nonsingular bounce by nonperturbative methods including numerical simulations of the nonsingular bounce and the covariant formalism for calculating nonlinear perturbations. We show that the bounce is disrupted in regions of the universe with significant inhomogeneity and anisotropy over the background energy density, but is achieved in regions that are relatively homogeneous and isotropic. Sufficiently small perturbations, consistent with observational constraints, can pass through the nonsingular bounce with negligible alteration from nonlinearity. We follow scale invariant perturbations generated in a matter-like contraction phase through the bounce. Their amplitude in the expansion phase is determined by the growing mode in the contraction phase, and the scale invariance is well preserved across the bounce.
\end{abstract}


\maketitle

\section{Introduction} \label{sec:introduction}

A bouncing cosmology is a scenario in which the universe transitions from a previous contraction phase to the current expansion phase through a ``big bounce'' \cite{Gasperini:2002bn, Wands:2008tv, Lehners:2011kr, Brandenberger:2012zb}. In this scenario, the primordial fluctuations that seed structure formation in the early expansion phase originate from adiabatic perturbations generated in the contraction phase. These adiabatic perturbations arise as quantum fluctuations when the modes are deep inside the horizon in the early contraction phase, then become classical when they exit the horizon during the contraction phase. A large number of nearly scale invariant modes can be produced by various mechanisms in the contraction phase \cite{Enqvist:2001zp, Lyth:2001nq, Wands:1998yp, Finelli:2001sr, Lehners:2007ac, Buchbinder:2007ad, Creminelli:2007aq, Khoury:2009my, Khoury:2011ii}. These modes then have to pass through the bounce and carry on to the expansion phase. Their power spectra would match current observations if the adiabatic perturbations remain nearly scale invariant after the bounce.

In a singular bounce, the universe passes through a classical singularity which must be resolved by a quantum theory of gravity. The prediction of such models relies on a quantum treatment of the singular bounce \cite{Gasperini:1992em, Gasperini:2002bn, Tolley:2003nx, Turok:2004gb, McFadden:2005mq}. Certain matching conditions can be derived from the requirement of analyticity and unitarity of the bouncing process \cite{Israel:1966rt, Hwang:1991an, Deruelle:1995kd, Durrer:2002jn, Finelli:2001sr, Bars:2011aa}. The alternative approach is a nonsingular bounce where the universe stops contraction and reverses to expansion at a finite size, which can be well described by classical general relativity and effective field theory \cite{Allen:2004vz, Buchbinder:2007ad, Creminelli:2007aq, Cai:2007qw, Lin:2010pf, Easson:2011zy, Cai:2012va}. In such a smooth transition the adiabatic perturbations evolve classically and can be followed directly through the bounce. A rigorous analysis of a nonsingular bounce would help display properties and build intuition about bouncing models, especially if the outcome of the bounce does not depend sensitively on the details of the bouncing mechanism.

However, there are several reasons to be concerned. First, a nonsingular bounce requires a violation of the null energy condition (NEC). In order for the cosmic contraction to slow down to a halt, the effective equation of state parameter of the universe, defined by $\frac{3}{2}(1+w) = -\dot{H}/H^2$, must fall below $w = -1$ for an extended period of time, referred to as the ``bouncing phase''. During this bouncing phase, curvature and anisotropy, with effective equations of state $w = -\frac{1}{3}$ and $1$ respectively, grow much faster than the background energy density. Such unstable growth can potentially lead the universe into chaotic mixmaster behavior that disrupts the bounce altogether \cite{Xue:2010ux, Xue:2011nw}. It is therefore important to determine if the growth of curvature and anisotropy can be kept at a finite level during the bounce.

Moreover, the growth of curvature perturbations during the bouncing phase may change the shape of the power spectrum \cite{Xue:2010ux}. The power spectrum is given by the amplitude of adiabatic perturbations of different wavenumbers. The adiabatic modes that exit the horizon in the contraction phase appear to reenter the horizon near the bounce, since the Hubble scale $1/aH$ diverges when $H \to 0$ at the nonsingular bounce whereas the scale factor $a$ remains finite. If the amplitude of the adiabatic perturbations grows during the bounce and the growth varies with wavenumber, then the power spectrum would be distorted away from scale invariance.

The rapid growth of perturbations also raises the question of whether their evolution becomes nonlinear during the bouncing phase. According to a simple estimate \cite{Leblond:2008gg, Baumann:2011dt}, the adiabatic perturbations become strongly coupled when the magnitude of the dimensionless parameter $\epsilon \equiv -\dot{H}/H^2$ is much larger than $1$. Towards a nonsingular bounce, $\epsilon$ approaches $-\infty$ because $H$ goes to zero and $\dot{H}$ is positive. If the strong coupling problem occurs, then cubic and higher order interaction terms in the action of the adiabatic perturbation become comparable to and even larger than the quadratic term. Accordingly, the classical evolution of superhorizon modes may deviate from linearized equations of motion. Such nonlinearity causes a mixing of modes that alters the power spectrum and induces a large non-Gaussianity. To analyze the nonlinear evolution of adiabatic perturbations, a nonperturbative calculation is required.

In this paper, we present a full numerical analysis of a nonsingular bounce. Classical evolution of the spacetime is followed from near the end of the contraction phase through the entire bouncing phase into the expansion phase. The Einstein equations coupled with equations of motion for scalar fields are solved with inhomogeneous and anisotropic initial conditions. These inhomogeneities represent adiabatic perturbations that have exited the horizon in the earlier contraction phase and henceforth evolved classically. Our purpose is to study their behavior during the nonsingular bounce.

Several mechanisms for creating a nonsingular bounce have been studied, which are based on effective field theories such as ghost condensation \cite{ArkaniHamed:2003uy, Buchbinder:2007ad, Creminelli:2007aq, Lin:2010pf} and the galileon \cite{Nicolis:2008in, Qiu:2011cy, Easson:2011zy, Cai:2012va}. These mechanisms are not well-adapted for numerical simulations because the covariant generalization of the effective field theories to curved spacetime typically introduce higher derivative terms that are susceptible to numerical instability \cite{Anisimov:2004sp, Deffayet:2009wt, Deffayet:2010qz}. For our computation we introduce a minimally coupled massless scalar field with a wrong-signed kinetic term -- a ghost field -- whose stress-energy tensor explicitly breaks the NEC. Such a ghost field would lead to unstable excitations of negative quanta when coupled to normal scalar fields and gravity. Here we treat it purely classically as an effective way of creating a background solution that describes a nonsingular bounce; its energy density is only significant near the bounce and otherwise negligible. The classical equation of motion for the ghost field has an identical form to that of a normal scalar field and is well-behaved for numerical computations. A similar setup has been used in the perturbative calculations of \cite{Peter:2002cn, Allen:2004vz, Cai:2007zv}.

In our nonperturbative calculation, the evolution of the spacetime through the bouncing phase is computed by using harmonic coordinates \cite{Garfinkle:2001ni, Pretorius:2004jg, Lindblom:2005qh}. Compared to other numerical schemes, like the constant mean curvature gauge, the harmonic gauge does not run into coordinate singularities during the bouncing phase when $a$ and $H$ are non-monotonic in time. Another advantage of using harmonic coordinates is that the equations of motion for metric components are wave-like equations that can be easily solved. To extract the amplitude of adiabatic perturbations from our numerical results, we compute the nonlinear and covariant generalization of the curvature perturbation used in linear perturbation theory. Following the covariant formalism \cite{Langlois:2005qp, Langlois:2006vv}, the generalized curvature perturbation is taken to be the integrated expansion along the integral curve of the normal vector to the constant time slices. Our numerical methods are sufficiently general and robust to handle nonlinear evolution with large inhomogeneities.

We show that inhomogeneity and anisotropy can disrupt the nonsingular bounce. In particular, if the effective density of anisotropy in certain regions of the universe surpasses the energy density of the ghost field that is responsible for the bounce, then these regions will keep contracting and collapse to a singularity. On the other hand, regions of the universe that are relatively homogeneous and isotropic (e.g., those that underwent an ekpyrotic smoothing phase prior to the bounce) can undergo a nonsingular bounce and continue into the expansion phase. This picture is dramatically different from that obtained in linear perturbation analysis where the bounce happens everywhere and almost simultaneously.

For sufficiently small perturbations that pass through the nonsingular bounce, we study the effect of nonlinearity by measuring the mixing of Fourier modes in the integrated expansion. We show that, if the amount of inhomogeneity and anisotropy is marginally below the level that would disrupt the bounce, then the nonlinear terms introduce significant deviation from linear perturbative calculations. Otherwise, if the amount of inhomogeneity and anisotropy is well below the critical level, then nonlinearity is negligible throughout the bounce despite the fact that $\epsilon \to -\infty$ at the bounce. We compare our results with the condition of strong coupling, and argue that the latter problem does not occur at the classical level.

We further examine how the bounce affects the shape of the power spectrum. For the purpose of illustration, we assume that scale invariant perturbations are generated from a matter-like contraction phase before the bounce \cite{Wands:1998yp, Finelli:2001sr}. We find the matching condition that the amplitude of the adiabatic perturbations in the expansion phase is dominated by the contribution from the growing mode in the contraction phase. The change in the scale dependence of the amplitude through the bounce is negligible for small perturbations. The power spectrum of the adiabatic perturbations remains to be scale invariant after the bounce with no observable tilt or running.

The nonsingular bouncing model is presented in Section~\ref{sec:bouncing}. The numerical methods for simulating the nonsingular bounce and computing nonlinear perturbations are explained in Section~\ref{sec:methods}. The results of large inhomogeneity and anisotropy are described in Section~\ref{sec:anisotropy}, the problem of nonlinearity and strong coupling is characterized in Section~\ref{sec:nonlinearity}, and the effect of the bounce on the scale dependence of the power spectrum is studied in Section~\ref{sec:spectrum}. For comparison, a perturbative calculation in the linear harmonic gauge is presented in Appendix~\ref{sec:linear}, with Bunch-Davies initial values given in Appendix~\ref{sec:initial}. Conclusions and implications are discussed in Section~\ref{sec:conclusion}.

\section{Nonsingular bouncing model} \label{sec:bouncing}

For the nonsingular bounce, we consider a model with two scalar fields $\phi$ and $\chi$ minimally coupled to gravity, described by the Lagrangian
\begin{equation} \label{eq:lagrangian}
\mathcal{L} = - \tfrac{1}{2} (\partial \phi)^2 - V(\phi) + \tfrac{1}{2} (\partial \chi)^2 .
\end{equation}
Here $\phi$ is a canonical scalar field with a potential $V(\phi) = V_0 \, e^{- c \phi}$, and $\chi$ is a ghost field with a wrong-signed kinetic term. The conditions are chosen so that the universe is dominated by the normal scalar field $\phi$ during the contraction phase. Under such conditions, the $\phi$ field has a scaling solution in which its energy density scales as $1/a^{3(1 + w_\phi)}$ with a constant equation of state $w_\phi = \frac{c^2}{3} - 1$. For $c > \sqrt{6}$ and $V_0 < 0$, this solution is an attractor with $w_\phi > 1$ that is used in the ekpyrotic model \cite{Khoury:2001wf, Erickson:2003zm, Garfinkle:2008ei}. Here we consider the other case with $c < \sqrt{6}$ and $V_0 > 0$, so that $w_\phi < 1$ and a nonsingular bounce can be obtained \cite{Allen:2004vz}. In this case, the scaling solution is \emph{not} an attractor -- the initial condition must be fine-tuned in order to keep $w_\phi$ nearly constant for a sustained period.

Our computation starts near the end of the contraction phase, assuming that the $\phi$ field has $w_\phi$ given by the scaling solution and the $\chi$ field has negligible energy density. Since the $\chi$ field has only a kinetic term, its equation of state is $w_\chi = 1$ that is greater than $w_\phi$. Therefore the negative energy density of the $\chi$ field grows as $1/a^6$ during contraction, faster than the positive energy density of the $\phi$ field. Eventually the total energy density vanishes and causes a nonsingular bounce, in which the contraction stops and the expansion begins. Then the energy density of the $\chi$ field quickly diminishes and becomes negligible again compared to that of the $\phi$ field.

In a homogeneous, flat, and isotropic background, the equations of motion for the scalar fields are
\begin{align}
\phi'' + a^6 V_{,\phi} &= 0 , \label{eq:harmhomophi} \\[4pt]
\chi'' &= 0 . \label{eq:harmhomochi}
\end{align}
Here $'$ denotes derivative with respect to the \emph{harmonic time} $t$, related to the physical time $\tau$ by $d\tau = a^3 dt$; it is chosen to satisfy the gauge condition (\ref{eq:harmonic}), as introduced in Section~\ref{sec:harmonic} below. The Friedmann equations in harmonic time are given by
\begin{align}
\mathcal{H}^2 &= \tfrac{1}{3} \big( \tfrac{1}{2} {\phi'}^2 + a^6 V - \tfrac{1}{2} {\chi'}^2 \big) , \label{eq:harmFried1} \\[4pt]
\mathcal{H}' &= a^6 V , \label{eq:harmFried2}
\end{align}
where the \emph{harmonic Hubble parameter} $\mathcal{H}$ is defined as $\mathcal{H} \equiv a' / a$.

\begin{figure}
\centering
\includegraphics[width=0.5\textwidth]{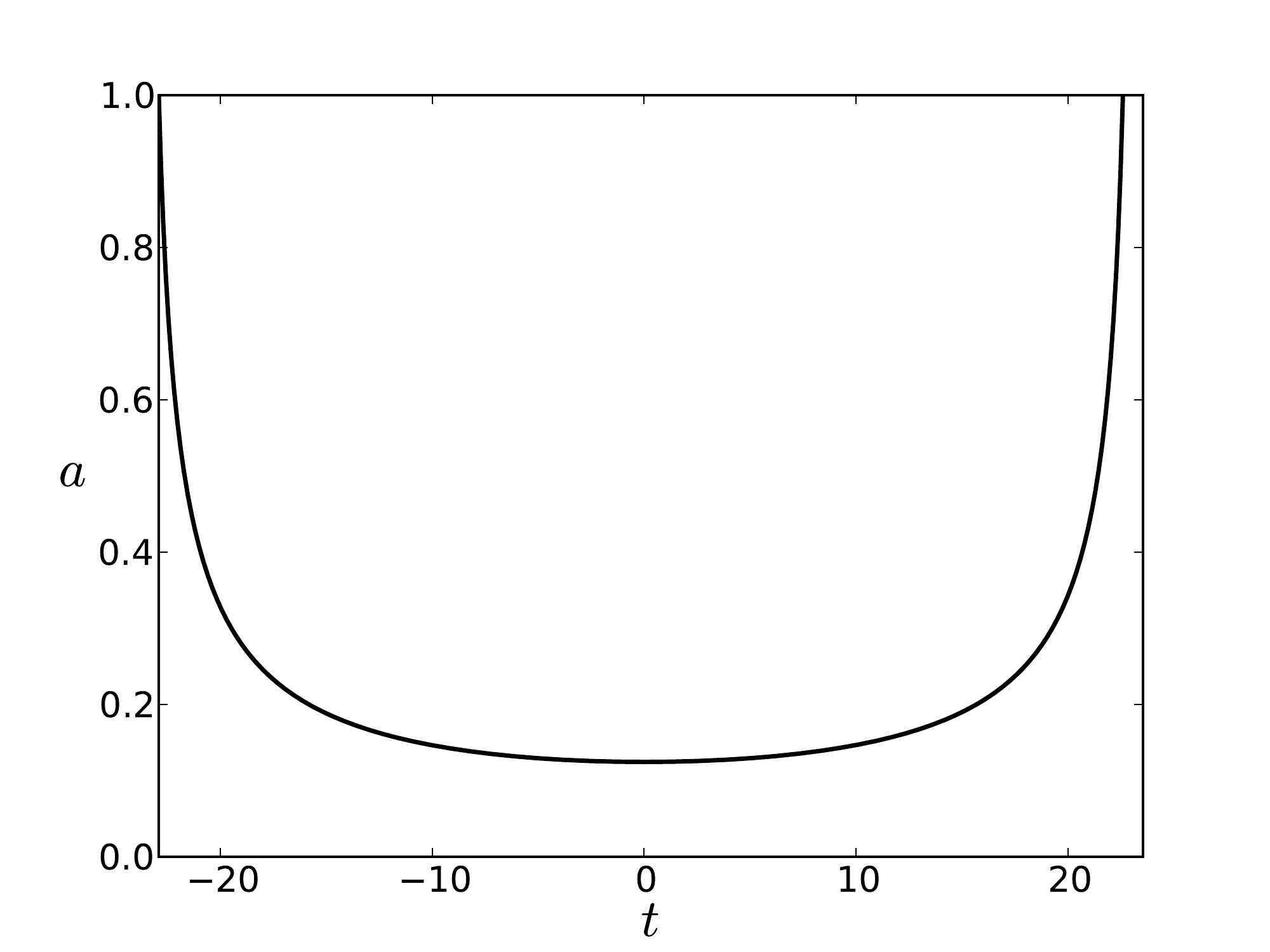}
\caption{The background solution of the scale factor $a$. The harmonic time $t$ is shifted so that the bounce occurs at $t=0$, and rescaled in units of $1/|H_0|$ where $H_0$ is the initial value of the Hubble parameter.} \label{fig:a}
\end{figure}

\begin{figure}
\centering
\includegraphics[width=0.5\textwidth]{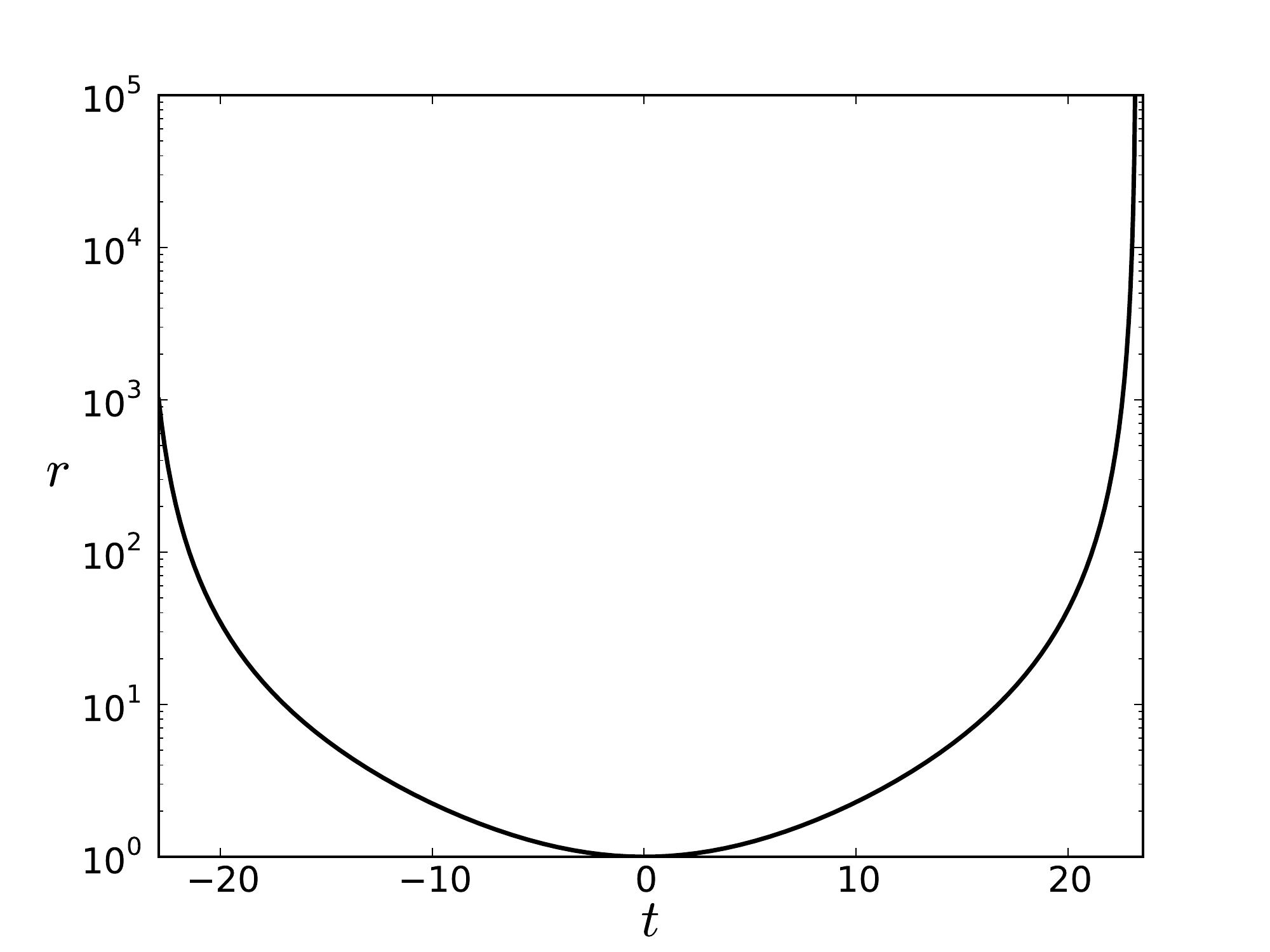}
\caption{The ratio between the energy density of the scalar fields $\phi$ and $\chi$. The $\chi$ field energy density is significant only near the bounce and otherwise negligible. The time coordinate is scaled in the same way as in Fig.~\ref{fig:a}.} \label{fig:r}
\end{figure}

The background solution can be found by evolving Eqs.~(\ref{eq:harmhomophi}, \ref{eq:harmhomochi}, \ref{eq:harmFried2}) and using (\ref{eq:harmFried1}) as a constraint. The initial values for $\phi$, $\phi'$, $\chi$, $\chi'$, $a$, and $\mathcal{H}$ are set by
\begin{alignat}{2}
\phi(0) &\equiv \phi_0 = 0 , \quad & \phi'(0) &\equiv a_0^3 \dot{\phi}_0 = - a_0^3 \sqrt{\frac{2 c^2 V_0}{6 - c^2}} , \label{eq:phi0} \\[4pt]
\chi(0) &\equiv \chi_0 = 0 , \quad & \chi'(0) &\equiv a_0^3 \dot{\chi}_0 = a_0^3 \sqrt{\frac{12 V_0}{(6 - c^2) r_0}} , \label{eq:chi0} \\[4pt]
a(0) &\equiv a_0 = 1 , \quad & \mathcal{H}(0) &\equiv a_0^3 H_0 = - a_0^3 \sqrt{\frac{2 V_0 (r_0 - 1)}{(6 - c^2) r_0}} , \label{eq:H0}
\end{alignat}
We choose $c = \sqrt{3}$ so that initially the $\phi$ field obeys the scaling solution with a matter-like equation of state, $w_\phi = 0$; such a matter-like contraction phase can generate scale invariant adiabatic perturbations before the bounce. $r_0$ represents the initial value of the ratio between the energy density of the $\phi$ field and the $\chi$ field, $|\rho_\phi / \rho_\chi|$. For $V_0 = 0.1$ and $r_0 = 1000$, the bouncing solution for the scale factor $a$ is shown in Fig.~\ref{fig:a}.
The ratio between the energy density of the $\phi$ and $\chi$ fields is shown in Fig.~\ref{fig:r},
illustrating that the $\chi$ field is only significant near the bounce.

The ghost field $\chi$ must be stabilized by some mechanism at the quantum level, which will not be considered in this paper. Here we only use its classical equation of motion to effectively describe the nonsingular bouncing process. This simple setup allows us to study the classical evolution of adiabatic modes that have left the horizon during the contraction phase. In particular, we will follow the amplitude of the curvature perturbation through the nonsingular bounce.

\section{Numerical methods} \label{sec:methods}

To analyze the nonsingular bouncing model nonperturbatively, we employ numerical methods to solve the equations for the spacetime metric and the scalar fields. Our methods allow a wide range of inhomogeneous, nonflat, and anisotropic initial conditions.

\subsection{Harmonic coordinates} \label{sec:harmonic}

The spacetime can be described by a coordinate system $(x^\mu) = (t, x^i)$ with a metric $g_{\mu\nu}$, where $t$ is a timelike coordinate and $x^i$ are spacelike coordinates. The full set of equations include the Einstein equations for the metric $g_{\mu\nu}$ and the equations of motion for the scalar fields $\phi$ and $\chi$. From the Lagrangian (\ref{eq:lagrangian}) it follows that the $\phi$ and $\chi$ fields satisfy the equations
\begin{align}
{\nabla_{\!\!\alpha}}{\nabla^\alpha} \phi &= V'(\phi) \, , \label{eq:eomphi} \\[4pt]
{\nabla_{\!\!\alpha}}{\nabla^\alpha} \chi &= 0 \, , \label{eq:eomchi}
\end{align}
where $\nabla_{\!\!\mu}$ denotes the covariant derivative associated with $g_{\mu\nu}$. The total stress-energy tensor is given by
\begin{equation}
{T_{\alpha \beta}} = {\nabla_{\!\!\alpha}}\phi \, {\nabla_{\!\!\beta}}\phi - {\nabla_{\!\!\alpha}}\chi \, {\nabla_{\!\!\beta}}\chi - {g_{\alpha \beta}}
\left( {\textstyle {\frac 1 2}} {\nabla^\gamma}\phi \, {\nabla_{\!\!\gamma}}\phi + V - {\textstyle {\frac 1 2}} {\nabla^\gamma}\chi \, {\nabla_{\!\!\gamma}}\chi \right) .
\end{equation}
Hence the Einstein equations can be written in the trace reversed form as
\begin{equation} \label{eq:einstein}
{R_{\alpha\beta}} = {\nabla_{\!\!\alpha}}\phi \, {\nabla_{\!\!\beta}}\phi - {\nabla_{\!\!\alpha}}\chi \, {\nabla_{\!\!\beta}}\chi + V {g_{\alpha\beta}} \, ,
\end{equation}
where we use reduced Planck units with $8 \pi G \equiv 1$.

To solve the Einstein equations, one must first remove the diffeomorphic freedom in the coordinates by fixing a gauge. This involves choosing a set of time slices, such as the constant mean curvature slices used in simulating the ekpyrotic contraction phase \cite{Garfinkle:2008ei}. However, the same method is not applicable to the bouncing phase. The mean curvature becomes non-monotonic in time when the universe enters the bouncing phase from the contraction phase, and when it exits the bouncing phase to enter the expansion phase. Consequently, in the presence of inhomogeneities, the constant mean curvature slices stop being spacelike during these transitions (see Section~\ref{sec:covariant} below), rendering the numerical evolution ill-behaved. So instead, we shall use a different gauge that is well-defined throughout the entire cosmic transition from contraction to expansion -- the \emph{harmonic gauge} \cite{Garfinkle:2001ni, Pretorius:2004jg, Lindblom:2005qh}.

The harmonic coordinates are defined to satisfy the gauge condition
\begin{equation} \label{eq:harmonic}
\nabla_{\!\!\alpha} \nabla^\alpha x^\mu = 0 \, .
\end{equation}
Consequently, the Christoffel symbols $\Gamma ^\gamma _{\ \alpha \beta }$ must satisfy the condition
\begin{equation} \label{eq:gGamma}
{g^{\alpha \beta}} {\Gamma ^\gamma _{\ \alpha \beta }} = 0 \, .
\end{equation}
Under this condition, the Ricci tensor takes the form
\begin{equation}
{R_{\alpha \beta }} = - {\textstyle {1 \over 2}} {g^{\gamma \sigma}} {\partial _\gamma} {\partial _\sigma} {g_{\alpha \beta}} + {g^{\lambda \mu}} {g^{\rho \nu}} {\partial _\mu} {g_{\nu (\alpha}} {\partial _{\beta)}} {g_{\lambda \rho}} - {\Gamma ^\gamma _{\ \sigma \alpha }} {\Gamma ^\sigma _{\ \gamma \beta }} \, .
\end{equation}
The first term controls the character of the equations, giving rise to hyperbolic differential equations for the metric components $g_{\alpha\beta}$. These $10$ equations are subject to the $4$ constraints given by (\ref{eq:gGamma}).

To solve the equations numerically, we first reduce them to first order differential equations in time. Define the variables $P_{\alpha \beta}$, $P_\phi$, and $P_\chi$ by
\begin{align}
{P_{\alpha \beta}} &\equiv {\partial_0} {g_{\alpha \beta}} \, ,\label{eq:pdef} \\[4pt]
{P_\phi} &\equiv {\partial_0} \phi \, , \label{eq:pphidef} \\[4pt]
{P_\chi} &\equiv {\partial_0} \chi \, . \label{eq:pchidef}
\end{align}
Then the Einstein equations (\ref{eq:einstein}) become
\begin{align} \label{eq:evolvep}
- {g^{00}} {\partial_0} {P_{\alpha \beta }} = &\ 2 {g^{0k}} {\partial _k} {P_{\alpha \beta}} + {g^{ik}} {\partial _i} {\partial _k} {g_{\alpha \beta}} - 2 {g^{\lambda \mu}} {g^{\rho \nu}} {\partial _\mu} {g_{\nu (\alpha}} {\partial _{\beta)}} {g_{\lambda \rho}} \nonumber \\
&+ 2 {\Gamma ^\gamma _{\ \sigma \alpha}} {\Gamma ^\sigma _{\ \gamma \beta }} + 2 \left( {\partial _\alpha } \phi {\partial _\beta} \phi - {\partial _\alpha } \chi {\partial _\beta} \chi + V {g_{\alpha \beta}} \right) \, ,
\end{align}
and the equations (\ref{eq:eomphi}, \ref{eq:eomchi}) for $\phi$ and $\chi$ become
\begin{align}
- {g^{00}} {\partial_0} {P_\phi} &= 2 {g^{0k}} {\partial _k} {P_\phi} + {g^{ik}} {\partial _i} {\partial _k} \phi - V'(\phi) \, , \label{eq:evolvepphi} \\[4pt]
- {g^{00}} {\partial_0} {P_\chi} &= 2 {g^{0k}} {\partial _k} {P_\chi} + {g^{ik}} {\partial _i} {\partial _k} \chi \, . \label{eq:evolvepchi}
\end{align}

To specify initial data, we choose the initial time slice to have constant mean curvature, $K = -3 H_0$. The full metric $g_{\mu\nu}$ can be decomposed as
\begin{equation} \label{eq:metric}
ds^2 = - \alpha^2 dt^2 + \gamma_{ij} (\beta^i dt + dx^i) (\beta^j dt + dx^j) \, ,
\end{equation}
where $\alpha$, $\beta^i$ are the lapse function and the shift vector, and $\gamma_{ij}$ is the spatial metric on the constant time slice. We can freely choose the lapse and the shift to be $\alpha = 1$ and $\beta^i = 0$ initially, then the spatial metric $\gamma_{ij}$ and its time derivative $\partial_0 \gamma_{ij} = -2 K_{ij}$ must satisfy the Hamiltonian and momentum constraints,
\begin{align}
{^{(3)}} \! R + {K^2} - {K^{ij}}{K_{ij}} &= \dot{\phi}^2 + {D^i} \phi \, {D_i} \phi + 2 V - \dot{\chi}^2 - {D^i} \chi \, {D_i} \chi \, , \label{eq:hamiltonian} \\[4pt]
{D_i} {{K^i}_j} - {D_j} K &= - \dot{\phi} \, {D_j} \phi + \dot{\chi} \, {D_j} \chi \, . \label{eq:momentum}
\end{align}
Here ${^{(3)}} \! R$ and $K_{ij}$ are the intrinsic and extrinsic curvature; $\dot{}$ denotes the derivative along the normal vector to the time slice, and $D_i$ is the covariant derivative associated with $\gamma_{ij}$. Once the above constraints are satisfied by the initial data, they will hold at all times as a result of the evolution equations and the harmonic coordinate condition \cite{Lindblom:2005qh}.

The constraint equations (\ref{eq:hamiltonian}, \ref{eq:momentum}) can be solved by using the York method \cite{York:1971hw, Gourgoulhon:2007tn}. Specifically, we choose the spatial metric to be conformally flat, and decompose the extrinsic curvature into the trace (i.e., mean curvature) and the traceless parts,
\begin{align}
\gamma_{ij} &\equiv \Psi ^4 \delta_{ij} \, , \\[4pt]
K_{ij} &\equiv \tfrac{1}{3} K \gamma_{ij} + \Psi^{-2} A_{ij} \, .
\end{align}
Define further the variables $Q_\phi$ and $Q_\chi$ by
\begin{align}
Q_\phi &\equiv {\Psi ^6} \dot{\phi} \, , \\[4pt]
Q_\chi &\equiv {\Psi ^6} \dot{\chi} \, .
\end{align}
The Hamiltonian and momentum constraints then become
\begin{align}
{\partial ^i}{\partial _i} \Psi &= - {\textstyle {\frac 1 8}} \big( {A^{ij}}{A_{ij}} + {Q_\phi}^2 - {Q_\chi}^2 \big) {\Psi ^{-7}} + {\textstyle {\frac 1 {12}}} ({K^2} - 3 V) {\Psi ^5} - {\textstyle {\frac 1 8}} \left ( {\partial ^i}\phi \, {\partial _i}\phi - {\partial ^i}\chi \, {\partial _i}\chi \right) \Psi \, , \label{eq:hamconstr} \\[4pt]
{\partial ^i}{A_{ij}} &= - {Q_\phi} {\partial_j}\phi + {Q_\chi} {\partial_j}\chi \, , \label{eq:momconstr}
\end{align}
where the indices in these two equations are raised and lowered with the flat metric $\delta_{ij}$.

For simplicity, we restrict our computation to the case with inhomogeneity only along one spatial dimension ($x$) with periodic boundary conditions. Then Eq.~(\ref{eq:momconstr}) is solved by the following ansatz,
\begin{align}
Q_\phi(x) &= \dot{\phi}_0 + f_0 \cos (m x) \, , \label{eq:f0} \\[4pt]
\phi(x) &= \phi_0 + f_1 \cos (m x) \, , \label{eq:f1} \\[4pt]
Q_\chi(x) &= \dot{\chi}_0 + f_2 \cos (m x) \, , \label{eq:f2} \\[4pt]
\chi(x) &= \chi_0 + f_3 \cos (m x) \, , \label{eq:f3}
\end{align}
and the particular solution
\begin{equation} \label{eq:Aij}
A_{ij}(x) = \left( \begin{array}{ccc} A_{11}(x) & 0 & 0 \\ 0 & \lambda A_{11}(x) & 0 \\ 0 & 0 & - (1 + \lambda) A_{11}(x) \end{array} \right) \, ,
\end{equation}
where $f_0$, $f_1$, $f_2$, $f_3$, and $\lambda$ are parameters to choose, and
\begin{equation} \label{eq:A11}
A_{11}(x) = - \dot{\phi}_0 f_1 \cos (m x) - \tfrac{1}{4} f_0 f_1 \cos (2 m x) + \dot{\chi}_0 f_3 \cos (m x) + \tfrac{1}{4} f_2 f_3 \cos (2 m x) \, .
\end{equation}
These expressions are then put into Eq.~(\ref{eq:hamconstr}) to solve for $\Psi(x)$, using a relaxation method. The results are substituted into the expressions for $\gamma_{ij}$ and its time derivative $P_{ij}$; the remaining components $g_{0\mu}$ are given by the lapse and the shift, and $P_{0\mu}$ are solved from the constraint (\ref{eq:gGamma}).

Thus our initial data are specified as follows:
\begin{align}
g_{00} (0,x) &= -1 \, , \label{eq:init_g00} \\[4pt]
g_{0i} (0,x) &= g_{i0} (0,x) = 0 \, , \label{eq:init_g0i} \\[4pt]
g_{ij} (0,x) &= \Psi(x)^4 \delta_{ij} \, , \label{eq:init_gij} \\[4pt]
P_{00} (0,x) &= 2 K \, , \label{eq:init_P00} \\[4pt]
P_{0i} (0,x) &= P_{i0} (0,x) = - 2 \Psi(x)^{-1} \partial_i \Psi(x) \, , \label{eq:init_P0i} \\[4pt]
P_{ij} (0,x) &= - \tfrac{2}{3} K \Psi(x)^4 \delta_{ij} - 2 \Psi(x)^{-2} A_{ij}(x) \, , \label{eq:init_Pij} \\[4pt]
\phi (0,x) &= \phi(x) \, , \label{eq:init_phi} \\[4pt]
P_\phi (0,x) &= \Psi(x)^{-6} Q_\phi(x) \, , \label{eq:init_Pphi} \\[4pt]
\chi (0,x) &= \chi(x) \, , \label{eq:init_chi} \\[4pt]
P_\chi (0,x) &= \Psi(x)^{-6} Q_\chi(x) \, . \label{eq:init_Pchi}
\end{align}
The parameters $\phi_0$, $\dot{\phi}_0$, $\chi_0$, $\dot{\chi}_0$ in (\ref{eq:f0} - \ref{eq:A11}) and $K = -3 H_0$ are chosen to match the background values given in (\ref{eq:phi0} - \ref{eq:H0}), whereas the parameters $f_0$, $f_1$, $f_2$, $f_3$, and $\lambda$ will be set to incorporate different amounts of inhomogeneity in the initial data. In the limit of small inhomogeneities, our choice of initial data represents a single Fourier mode with comoving wavenumber $k = m$ (see Appendix~\ref{sec:linear}). Notice however the terms with double wavenumber $k = 2m$ in Eq.~(\ref{eq:A11}), which represent small nonlinearities that are second order in $f_i$.

The ansatz (\ref{eq:f0} - \ref{eq:A11}) can also be generalized to include multiple modes. As an illustration, the ansatz for two Fourier modes $k = m_1$ and $m_2$ is given by
\begin{align}
Q_\phi(x) &= \dot{\phi}_0 + f_0 \cos (m_1 x + d_1) + g_0 \cos (m_2 x + d_2) \, , \label{eq:g0} \\[4pt]
\phi(x) &= \phi_0 + f_1 \cos (m_1 x + d_1) + g_1 \cos (m_2 x + d_2) \, , \label{eq:g1} \\[4pt]
Q_\chi(x) &= \dot{\chi}_0 + f_2 \cos (m_1 x + d_1) + g_2 \cos (m_2 x + d_2) \, , \label{eq:g2} \\[4pt]
\chi(x) &= \chi_0 + f_3 \cos (m_1 x + d_1) + g_3 \cos (m_2 x + d_2) \, , \label{eq:g3}
\end{align}
and
\begin{align} \label{eq:A11g}
A_{11}(x) = &- \big( \dot{\phi}_0 f_1 - \dot{\chi}_0 f_3 \big) \cos (m_1 x + d_1) - \tfrac{1}{4} \big( f_0 f_1 - f_2 f_3 \big) \cos (2 m_1 x + 2 d_1) \nonumber \\*
&- \big( \dot{\phi}_0 g_1 - \dot{\chi}_0 g_3 \big) \cos (m_2 x + d_2) - \tfrac{1}{4} \big( g_0 g_1 - g_2 g_3 \big) \cos (2 m_2 x + 2 d_2) \nonumber \\*
&- \frac{(f_0 g_1 m_2 + f_1 g_0 m_1) - (f_2 g_3 m_2 + f_3 g_2 m_1)}{2 (m_1 + m_2)} \cos \big( (m_1 + m_2) x + (d_1 + d_2) \big) \nonumber \\*
&- \frac{(f_0 g_1 m_2 - f_1 g_0 m_1) - (f_2 g_3 m_2 - f_3 g_2 m_1)}{2 (m_2 - m_1)} \cos \big( (m_2 - m_1) x + (d_2 - d_1) \big) \, .
\end{align}
The parameters $f_0$, $f_1$, $f_2$, $f_3$ specify the Fourier mode $k = m_1$ as before, whereas the new parameters $g_0$, $g_1$, $g_2$, $g_3$ are chosen to specify the second mode with $k = m_2$. Notice the appearance of mixed modes with $k = m_2 \pm m_1$ in (\ref{eq:A11g}); their amplitude is quadratically suppressed initially, just like the double wavenumber modes with $k = 2 m_1$ and $2 m_2$.

In our numerical computation, starting from the initial values (\ref{eq:init_g00} - \ref{eq:init_Pchi}), Eqs.~(\ref{eq:pdef} - \ref{eq:evolvepchi}) are evolved until one of the grid points first reaches future infinity. This is possible because the physical time $\tau = +\infty$ is compactified to a finite harmonic time $t$; indeed, for a homogeneous expansion with $a \sim \tau^{2/3(1+w)}$ and $w < 1$, the integral $t = \int a^{-3} d\tau$ converges at $\tau = +\infty$. The dynamical equations are evolved by using the iterated Crank-Nicholson method, with spatial derivatives evaluated by using standard second-order-accurate centered finite differences. The numerical convergence is tested by repeating the computation at successively higher resolutions and computing the left hand side of the constraint equation (\ref{eq:gGamma}); the numerical residues vanish quadratically with the resolution, confirming second order convergence. The results presented below are computed at a baseline resolution with $128$ grid points, and a CFL factor of $0.5$. Typical errors of the numerical solutions calculated from convergence studies are less than $\sim 0.1\%$.

\subsection{Covariant formalism} \label{sec:covariant}

The above numerical scheme will be used to calculate the amplitude of adiabatic perturbations. The adiabatic modes are often studied by using the gauge invariant variable $\zeta$, defined as the curvature perturbation on the uniform density slicing \cite{Bardeen:1983qw}. This quantity is convenient for studying the power spectrum of primordial fluctuations in the expansion phase, because it is invariant under gauge transformations as well as conserved on superhorizon scales. However, it is defined as a particular combination of linear perturbations of the metric and scalar fields. To extract the amplitude from our numerical computations, we look for a nonlinear generalization of the curvature perturbation that can be covariantly defined.

Here we follow the covariant formalism \cite{Hawking:1966qi, Ellis:1989jt, Bruni:1991kb, Langlois:2005qp, Langlois:2006vv}. In this approach, the cosmological perturbations are defined in a geometrical way without referring to specific coordinates. Such covariant variables are interpreted as perturbations because they vanish identically in a homogeneous, flat, and isotropic background; but they are fully nonperturbative quantities not restricted to linear order in a perturbative expansion. Therefore it is appropriate to use such variables to study the nonlinear evolution of adiabatic perturbations. This approach is closely related to the $\delta N$ formalism \cite{Starobinsky:1986fxa, Salopek:1990jq, Sasaki:1995aw, Lyth:2004gb, Lyth:2005fi}. The latter approach is often used to calculate the superhorizon curvature perturbations based on the separate universe approximation \cite{Wands:2000dp, Rigopoulos:2003ak}, and can be shown to agree with the covariant formalism on large scales \cite{Suyama:2012wi}.

Let $n_\mu$ be the timelike normal vector to the constant time slices. This unit vector can be regarded as the 4-velocity of a fiducial Eulerian observer \cite{Gourgoulhon:2007ue}, for whom the constant time slice is truly synchronous. Therefore the spacetime as decomposed in a particular 3+1 slicing describes the cosmic evolution as measured by the corresponding Eulerian observer. The worldline of the Eulerian observer is the integral curve of the normal vector $n^\mu$. The volume expansion of the congruence of those worldlines is given by
\begin{equation}
\theta \equiv \nabla_{\!\!\mu} n^\mu \, ,
\end{equation}
which represents three times the local Hubble expansion rate. Then the integrated expansion can be defined as
\begin{equation}
\mathcal{N} \equiv \int \frac{\theta}{3} \, d\tau \, ,
\end{equation}
where the integration is along the integral curve of $n^\mu$, and $\tau$ is the proper time given by the lapse function $\alpha$ through $d\tau = \alpha \, dt$. The integrated expansion $\mathcal{N}$ can be considered as the local number of e-folds of Hubble expansion,
and is defined up to an integration constant for each worldline. It is a covariant quantity that satisfies the equation
\begin{equation} \label{eq:Ndot}
\dot{\mathcal{N}} = \frac{\theta}{3} \; , \quad \hbox{i.e.} \quad n^\mu \partial_\mu \mathcal{N} = \frac{1}{3} \, \nabla_{\!\!\mu} n^\mu \, .
\end{equation}
Note that this quantity $\mathcal{N}$ depends on the choice of the spacetime slicing through the normal vector $n^\mu$.

In the covariant formalism, the integrated expansion $\mathcal{N}$ is used to define a covector \cite{Langlois:2005qp}
\begin{equation}
\zeta_\mu \equiv \partial_\mu \mathcal{N} - \frac{\dot{\mathcal{N}}}{\dot{\rho}} \partial_\mu \rho \, ,
\end{equation}
whose components $\zeta_i$ describe the spatial gradient of $\mathcal{N}$ on the uniform density slice where $\rho =$ const. This covector $\zeta_\mu$ vanishes in a homogeneous background, and in that sense defines a true perturbation that is fully nonperturbative. It is a generalization of the linear perturbation
\begin{equation}
\zeta \equiv - \psi - \frac{H}{\dot{\rho}} \delta\rho \, ,
\end{equation}
which describes the curvature perturbation $\psi$ (see Appendix~\ref{sec:linear}) in the uniform density gauge where $\delta\rho = 0$.

More generally, the nonlinear curvature perturbation in a particular gauge with normal vector $n_\mu$ can be described by the covector
\begin{equation}
- \psi_\mu \equiv D_\mu \mathcal{N} = \partial_\mu \mathcal{N} - \dot{\mathcal{N}} n_\mu \, ,
\end{equation}
where the integrated expansion $\mathcal{N}$ is defined with respect to the same vector $n^\mu$. Here $D_\mu$ is the covariant derivative projected onto the spatial hypersurface, $D_\mu \equiv (g_{\mu\nu} + n_\mu n_\nu) \nabla^\nu$. Note that $- \psi_i = \partial_i \mathcal{N}$ in the coordinates adapted to the slicing, since the spatial components $n_i$ vanish identically. In general, at linear order, $\psi_i$ reduces to the gradient of the linear curvature perturbation $\psi$ in the same gauge \cite{Langlois:2005qp}. Therefore, the negative integrated expansion $- \mathcal{N}$ is a covariant and nonlinear generalization of the linear curvature perturbation $\psi$.

Indeed, the homogeneous part of $\mathcal{N}$ equals the number of e-folds $N$ in the homogeneous background,
\begin{equation} \label{eq:N0}
N = \int \mathcal{H} \, dt = \ln a \, ,
\end{equation}
where the scale factor $a$ is set to be $1$ initially. At linear order, the inhomogeneous part of $\mathcal{N}$ is given by, up to an integration constant, \cite{Langlois:2006vv}
\begin{equation} \label{eq:N1}
\mathcal{N}^{(1)} = - \psi + \frac{1}{3} \int \nabla^2 \sigma \, dt \, ,
\end{equation}
where $\sigma$ is the shear perturbation (see Appendix~\ref{sec:linear}). On superhorizon scales, neglecting the gradient term, the inhomogeneous part of $\mathcal{N}$ then becomes
\begin{equation} \label{eq:deltaN}
\delta N \equiv \mathcal{N} - N \approx - \psi \, ,
\end{equation}
provided that $\mathcal{N} = -\psi$ on the initial time slice. This is the $\delta N$ formula for computing the curvature perturbation $\psi$ on superhorizon scales \cite{Sasaki:1995aw, Lyth:2004gb}. In practice, $\mathcal{N}$ is often calculated by making the separate universe approximation that $\mathcal{N}(t,x^i) \approx N(\phi^I (t_0,x^i))$, where $N(\phi^I (t_0,x^i))$ is the homogeneous number of e-folds as a function of the scalar fields $\phi^I$ on different patches of the initial time slice \cite{Sasaki:1995aw, Lyth:2004gb, Lyth:2005fi}.

Instead of using the $\delta N$ formalism, we can solve for $\mathcal{N}$ directly from Eq.~(\ref{eq:Ndot}). To calculate the perturbation in a particular gauge, $n_\mu$ should be chosen as the unit normal vector to the corresponding time slices. In harmonic coordinates, the normal vector to the constant harmonic time slices is given by
\begin{equation}
n^\text{(h)}_\mu = \big( - 1 / \sqrt{-g^{00}} , 0, 0, 0 \big) .
\end{equation}
This vector will be used in Eq.~(\ref{eq:Ndot}) to compute $\mathcal{N}^\text{(h)}$ in the harmonic gauge. The initial value is set to be $\mathcal{N}^\text{(h)} (0,x) = 2 \log \Psi(x)$, so that $g_{ij}(0,x) = e^{2 \mathcal{N}(0,x)} \delta_{ij}$ by Eq.~(\ref{eq:init_gij}).

Ideally, we would also like to calculate the integrated expansion $\mathcal{N}^{(\phi)}$ on the constant $\phi$ hypersurface, which is the generalization of the curvature perturbation $-\mathcal{R}_\phi$ in the comoving $\phi$ gauge. $\mathcal{N}^{(\phi)}$ should be calculated by using $n^{(\phi)}_\mu = \partial_\mu \phi / \sqrt{-(\partial \phi)^2}$, which is the normal vector to the constant $\phi$ hypersurfaces. However, the comoving $\phi$ gauge is not well-defined in the bouncing phase when there is inhomogeneity. Near the bounce the $\phi$ field switches from decreasing to increasing, causing $\partial_0 \phi$ to vanish at a certain point along the worldline. As a result, the normal vector $n^{(\phi)}_\mu$ stops being timelike near that point, and the constant $\phi$ hypersurface fails to be a spatial slicing. (The same problem happens in other commonly used gauges as well, including the uniform density gauge, constant mean curvature gauge, and uniform integrated expansion gauge.) Therefore we cannot calculate $\mathcal{N}^{(\phi)}$ directly in the numerical computation.

Hence, in our numerical computations, we will first solve for $\mathcal{N}^\text{(h)}$ in the harmonic gauge. Our purpose is to check whether its evolution becomes nonlinear, in which case the adiabatic perturbations would no longer remain scale invariant after the bounce. Otherwise, if nonlinearity remains small during the bouncing phase, then the curvature perturbation can be reliably calculated by linear perturbation theory. In that case, the comoving curvature perturbation $\mathcal{R}_\phi$ can be reconstructed from our results in the harmonic gauge by
\begin{equation} \label{eq:reconstr}
\mathcal{R}_\phi \approx - \delta N + \frac{\dot{N}}{\dot{\phi}^{(0)}} \, \delta\phi \, ,
\end{equation}
where $\phi^{(0)}$ and $\delta\phi$ are the homogeneous part of $\phi$ and the deviation from it, similar to Eq.~(\ref{eq:deltaN}). We will use the reconstructed amplitude of $\mathcal{R}_\phi$ to study the power spectrum of the adiabatic perturbations.

\section{Results} \label{sec:results}

We explore three questions concerning the evolution of adiabatic perturbations through the nonsingular bounce. First, can inhomogeneity and anisotropy ever grow enough to disrupt the bounce altogether? Second, if the perturbations can pass through the bounce without disrupting it, does their evolution become sufficiently nonlinear or strongly coupled to cause mode-mixing and distortion? Third, even if the nonlinearity is negligible throughout the bounce, does the scale dependence of the amplitude change as a result of the bounce so as to tilt the power spectrum away from scale-invariance? We show that for sufficiently small perturbations, consistent with the observed amplitude of primordial fluctuations, the answer to all three questions are negative.

\subsection{Inhomogeneity and anisotropy} \label{sec:anisotropy}

Our simulation of the nonsingular bounce starts from the initial data given in (\ref{eq:init_g00} - \ref{eq:init_Pchi}), with the ansatz (\ref{eq:f0} - \ref{eq:A11}) that describes a single Fourier mode in the limit of small perturbations. In addition to the dynamical variables $\phi$, $\chi$, and $g_{\mu\nu}$, we calculate the volume expansion $\theta$ at every spatial point. The Hamiltonian constraint (\ref{eq:hamconstr}) can be written as a generalized Friedmann equation \cite{Xue:2011nw},
\begin{equation} \label{eq:localFriedmann}
(\tfrac{1}{3} \theta)^2 = \tfrac{1}{3} \big( E_\phi + E_\chi - \tfrac{1}{2} {}^{(3)}\!R + \sigma^2 \big) .
\end{equation}
Here $E_\phi$ and $E_\chi$ are the energy density of the scalar fields,
\begin{align}
E_\phi &= \tfrac{1}{2} \big( \dot{\phi}^2 + {D^i} \phi \, {D_i} \phi \big) + V(\phi) \, , \\[4pt]
E_\chi &= - \tfrac{1}{2} \big( \dot{\chi}^2 + {D^i} \chi \, {D_i} \chi \big) \, ,
\end{align}
${}^{(3)}\!R$ is the spatial curvature, and $\sigma^2$ measures the amount of anisotropy,
\begin{equation}
\sigma^2 \equiv \tfrac{1}{2} \sigma^{ij} \sigma_{ij} \equiv \tfrac{1}{2} \big( K^{ij} - \tfrac{1}{3} K \delta^{ij} \big) \big( K_{ij} - \tfrac{1}{3} K \delta_{ij} \big) \, .
\end{equation}
Our computational results are presented in terms of the expansion $\theta$, normalized by $3 |H_0|$ where $H_0$ is the initial value of the Hubble parameter in the homogeneous solution, as well as the ratios $|E_\phi / E_\chi|$, $|\tfrac{1}{2} {}^{(3)}\!R / E_\chi|$, and $|\sigma^2 / E_\chi|$, where the last ratio represents the relative amount of anisotropy as compared to the $\chi$ field energy density.

\begin{figure}
\centering
\includegraphics[width=0.5\textwidth]{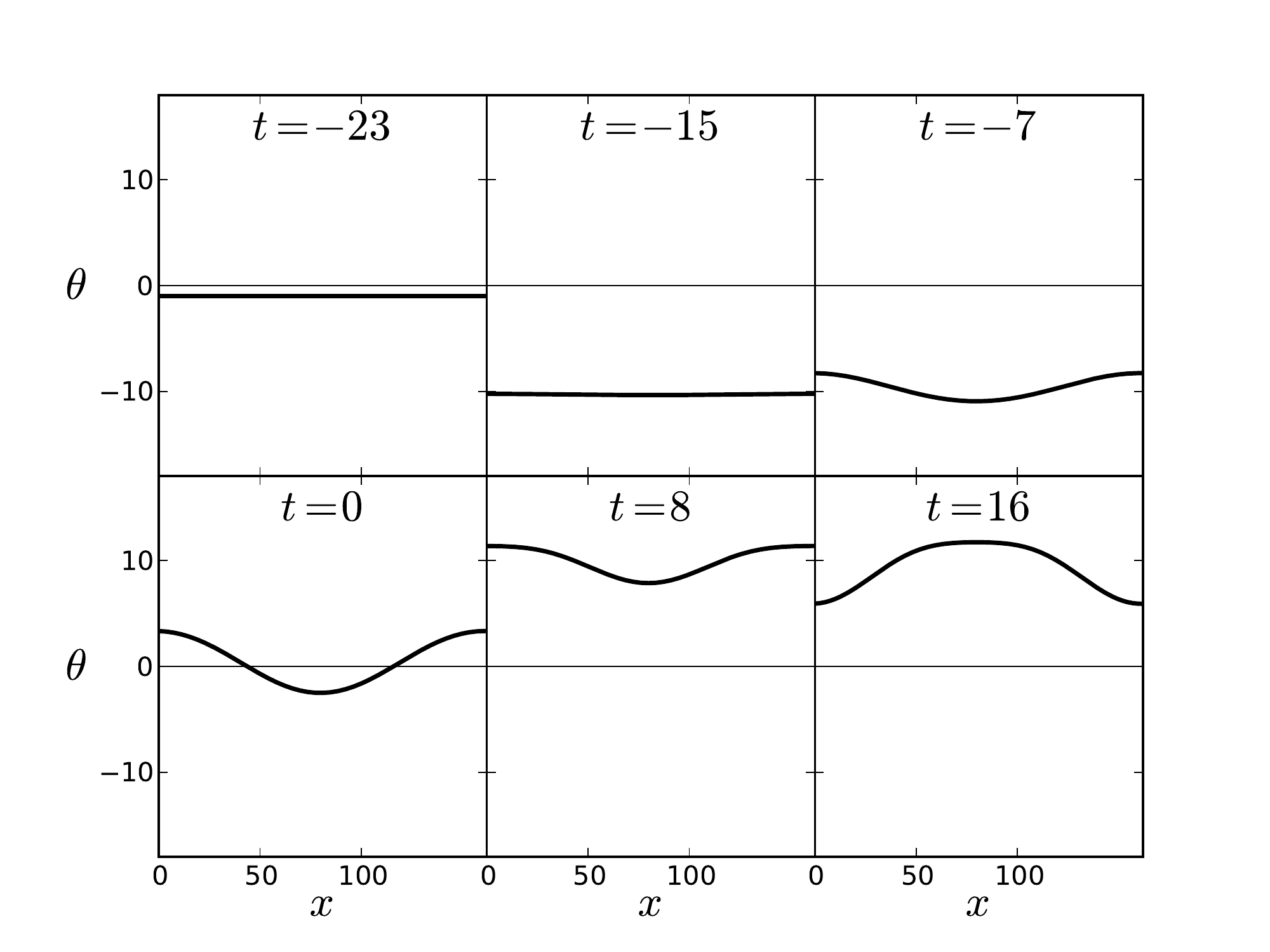}
\caption{Local expansion $\theta$ as a function of the coordinate $x$ at select times, computed with parameters in (\ref{eq:amp-3}). The spatial coordinate $x$ is scaled in units of $1/|H_0|$ which is the size of a Hubble length at the initial time, $t$ is scaled in the same way as in Fig.~\ref{fig:a} so that $t=0$ corresponds to the bounce in the homogeneous solution. The expansion $\theta$ is scaled in units of $3|H_0|$. In this inhomogeneous case, the nonsingular bounce occurs when $\theta$ crosses zero from below, which happens at different times for different spatial points.} \label{fig:theta-3}
\end{figure}

\begin{figure}
\centering
\includegraphics[width=0.5\textwidth]{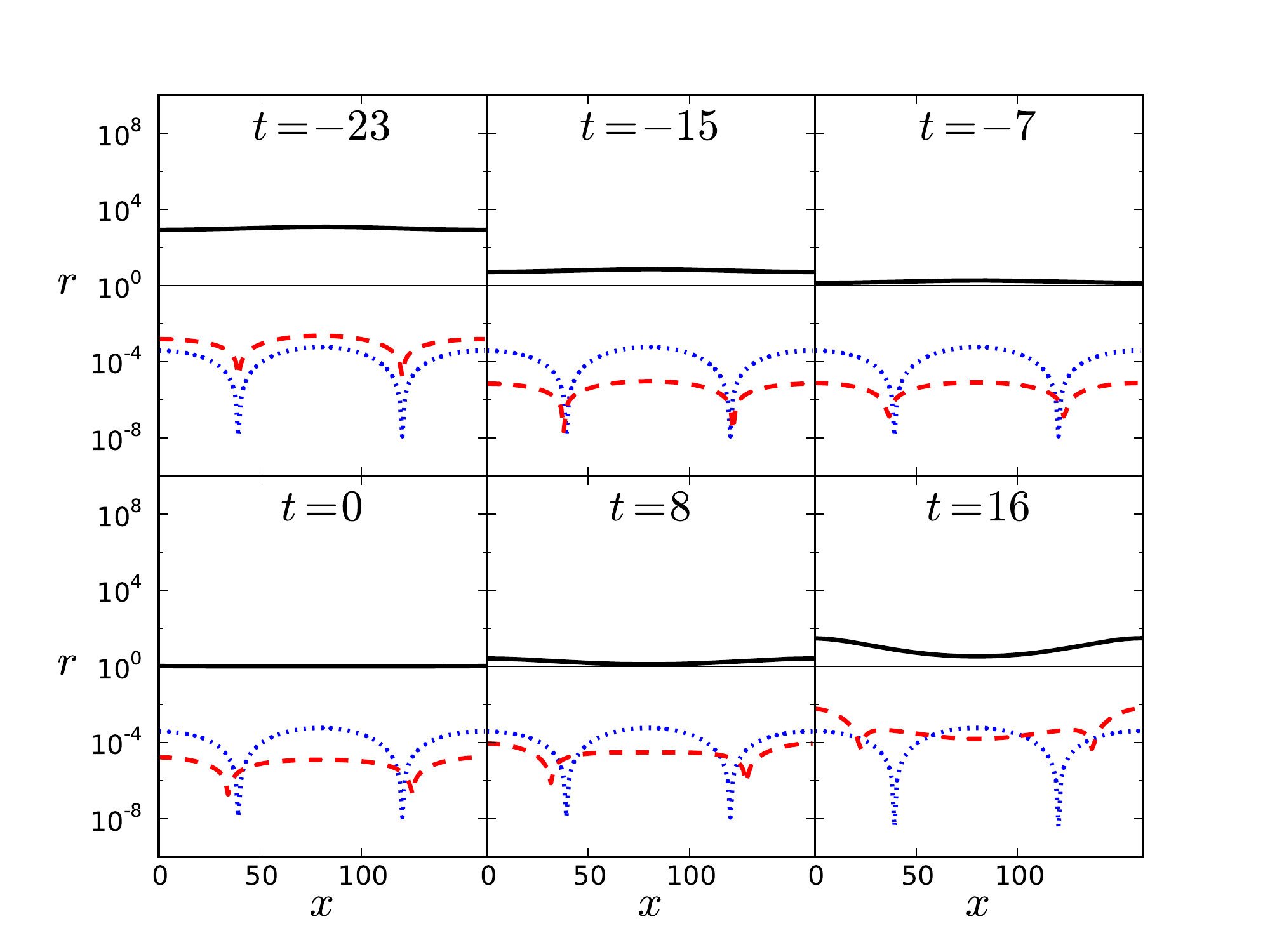}
\caption{(color online) $|E_\phi / E_\chi|$ (black continuous), $|\tfrac{1}{2} {}^{(3)}\!R / E_\chi|$ (red dashed), and $|\sigma^2 / E_\chi|$ (blue dotted) as a function of the coordinate $x$ at select times, computed with parameters in (\ref{eq:amp-3}). $x$ and $t$ coordinates are scaled in the same way as in Fig.~\ref{fig:theta-3}. In this example, curvature and anisotropy are negligible compared to the energy density of the scalar fields.} \label{fig:rho-3}
\end{figure}

Here is an example in which the universe undergoes a smooth nonsingular bounce, with parameters
\begin{alignat}{2} \label{eq:amp-3}
m &= 0.01 , \quad & \lambda &= -0.3 . \nonumber \\
f_0 &= -0.003 , \quad & f_1 &= 0.001 , \\
f_2 &= 0.002 , \quad & f_3 &= -0.005 . \nonumber
\end{alignat}
The numerical result for the expansion $\theta$ at select times is plotted in Fig.~\ref{fig:theta-3}.
The nonsingular bounce happens when the expansion $\theta$ crosses zero from below. Note that, due to inhomogeneities, the bounce happens at different times for different spatial points. Fig.~\ref{fig:rho-3} shows the ratios $|E_\phi / E_\chi|$, $|\tfrac{1}{2} {}^{(3)}\!R / E_\chi|$, and $|\sigma^2 / E_\chi|$, as defined in Eq.~(\ref{eq:localFriedmann}).
It can be seen that the magnitude of curvature and anisotropy remain small compared to the energy density of the scalar fields. Note that the ratio $|\sigma^2 / E_\chi|$ stays constant over time, just like in the homogeneous case where they have the same (effective) equation of state $w = 1$. The ratio $|E_\phi / E_\chi|$ shows that the $\chi$ field energy density starts much smaller than that of the $\phi$ field and becomes substantial only near the bounce, as in the homogeneous case shown in Fig.~\ref{fig:r}.

\begin{figure}
\centering
\includegraphics[width=0.5\textwidth]{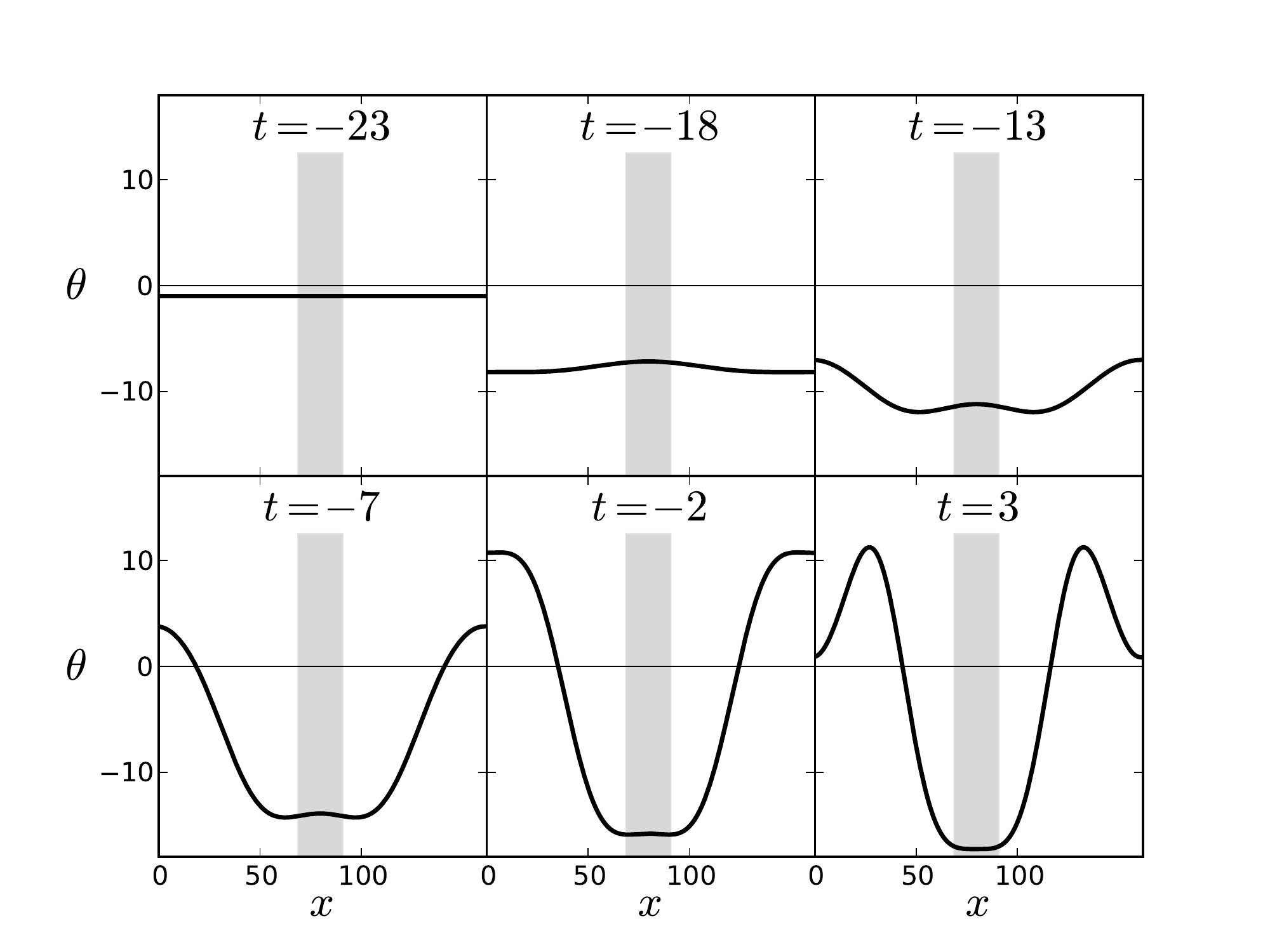}
\caption{Local expansion $\theta$ as a function of the coordinate $x$ at select times, computed with parameters in (\ref{eq:amp-2}). $x$ and $t$ coordinates and the expansion $\theta$ are scaled in the same way as in Fig.~\ref{fig:theta-3}. The nonsingular bounce does not occur in the shaded region (shown here in the middle of the $x$ range, though recall that $x$ is periodic and each panel covers a single period) where the ratio $|\sigma^2 / E_\chi|$ is greater than $1$, see Fig.~\ref{fig:rho-2}.} \label{fig:theta-2}
\end{figure}

\begin{figure}
\centering
\includegraphics[width=0.5\textwidth]{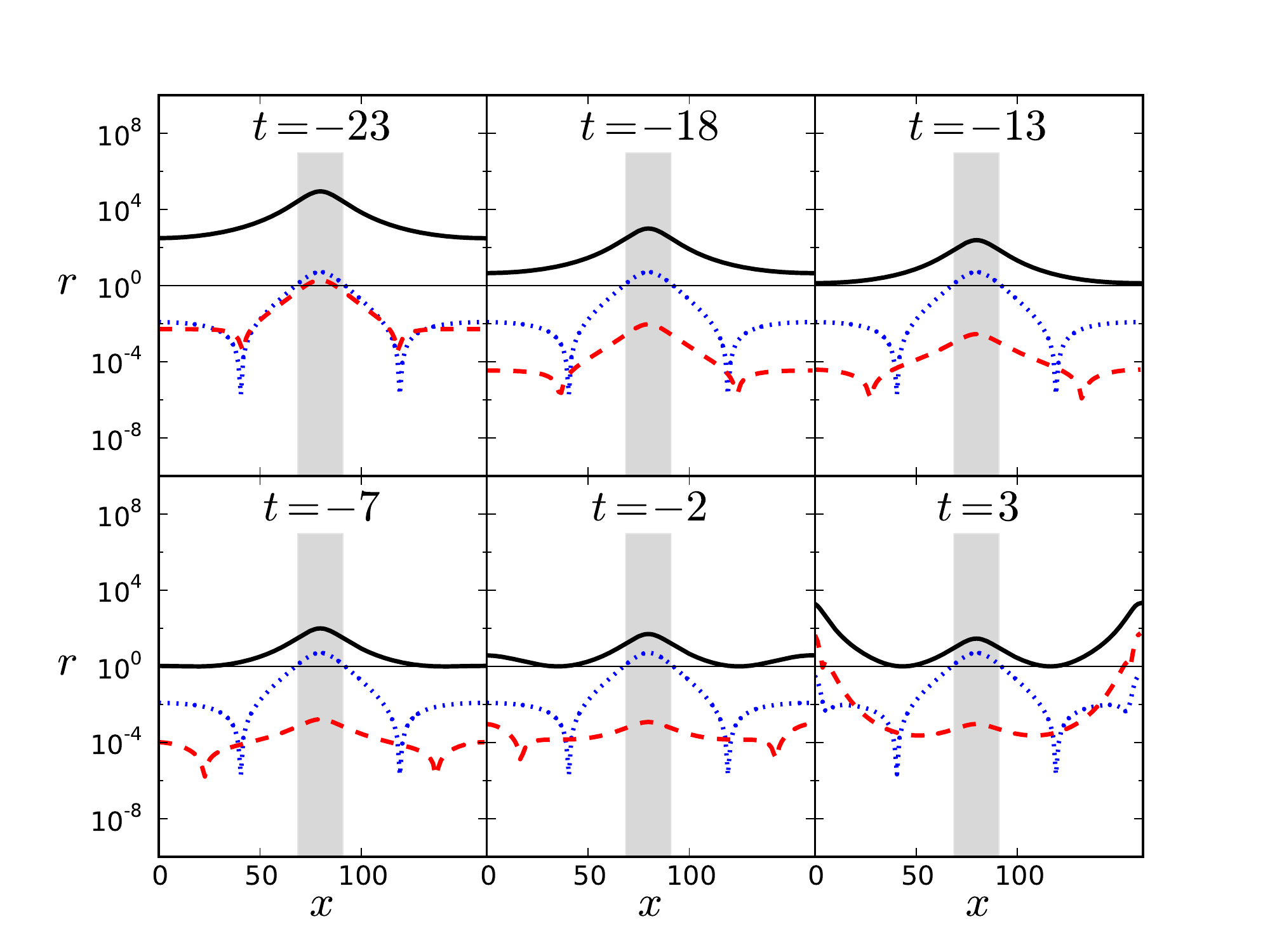}
\caption{(color online) $|E_\phi / E_\chi|$ (black continuous), $|\tfrac{1}{2} {}^{(3)}\!R / E_\chi|$ (red dashed), and $|\sigma^2 / E_\chi|$ (blue dotted) as a function of the coordinate $x$ at select times, computed with parameters in (\ref{eq:amp-2}); see Fig.~\ref{fig:theta-2} for the expansion at corresponding times. $x$ and $t$ coordinates are scaled in the same way as in Fig.~\ref{fig:theta-3}. In this example, $|\sigma^2 / E_\chi| > 1$ in the shaded region (shown here in the middle of the $x$ range), preventing a nonsingular bounce.} \label{fig:rho-2}
\end{figure}

Consider another example in which the initial perturbation amplitude is larger than the amplitude in (\ref{eq:amp-3}), with parameters
\begin{alignat}{2} \label{eq:amp-2}
f_0 &= -0.03 , \quad & f_1 &= 0.01 , \nonumber \\
f_2 &= 0.018 , \quad & f_3 &= -0.05 .
\end{alignat}
In this example, the expansion $\theta$ remains negative in the middle range of the coordinate $x$ shown in Fig.~\ref{fig:theta-2},
indicating that this part of the universe keeps contracting and never bounces. The reason is that, in this region the negative energy density of the $\chi$ field, which is supposed to induce the bounce, is overtaken by the anisotropy. As shown in Fig.~\ref{fig:rho-2},
the ratio $|\sigma^2 / E_\chi|$ is greater than $1$ in the shaded region. Since anisotropy grows at the same rate as the $\chi$ field energy density, the latter will never overtake the anisotropy to induce the bounce. Hence this part of the universe will collapse into a singularity, in contrast to the rest of space that will pass through a nonsingular bounce.

This example presents a scenario of nonsingular bouncing cosmology in which the nonsingular bounce does not occur everywhere in the universe, but only in separate regions that are relatively homogeneous and isotropic. The difference in the future evolution of separate regions is caused by large inhomogeneities that can only be precisely calculated using a nonperturbative approach such as the one presented here.

A quantitative figure-of-merit for determining whether a certain part of the universe will undergo a nonsingular bounce is the ratio $|\sigma^2 / E_\chi|$ between the anisotropy and the energy density of the $\chi$ field. Since this ratio remains constant during the cosmic evolution in our model, it is already set by the initial data. Therefore, regions where this ratio is initially less than $1$ will undergo a nonsingular bounce, whereas regions with a ratio greater than $1$ will not.

\begin{figure}
\centering
\includegraphics[width=0.5\textwidth]{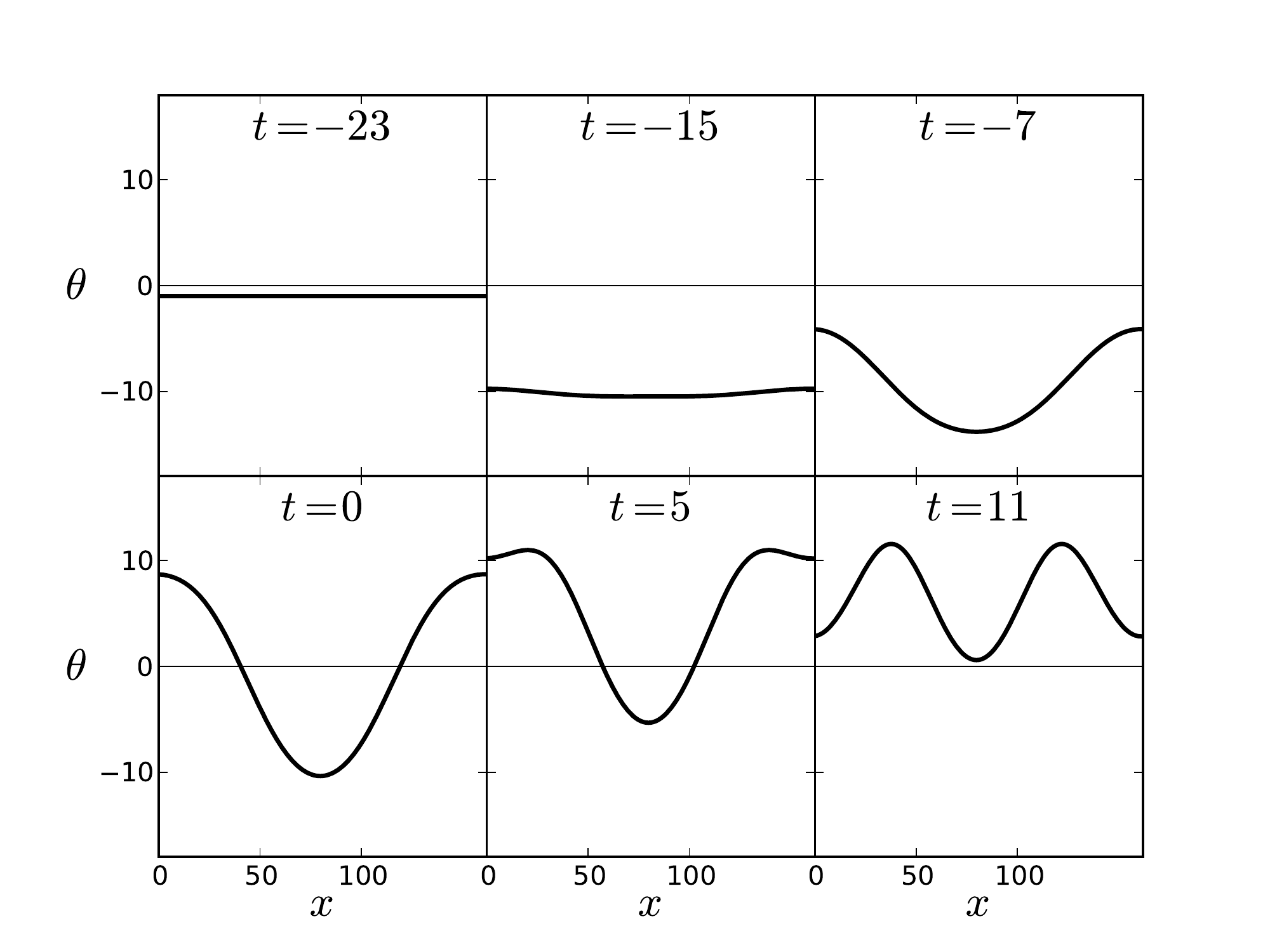}
\caption{Local expansion $\theta$ as a function of the coordinate $x$ at select times, computed with parameters in (\ref{eq:amp-2.5}). $x$ and $t$ coordinates and the expansion $\theta$ are scaled in the same way as in Fig.~\ref{fig:theta-3}. Part of the space (shown here in the middle region of the range of the periodic $x$ coordinate) bounces at a much later time compared to the other regions, causing inhomogeneity at late times.} \label{fig:theta-2.5}
\end{figure}

\begin{figure}
\centering
\includegraphics[width=0.5\textwidth]{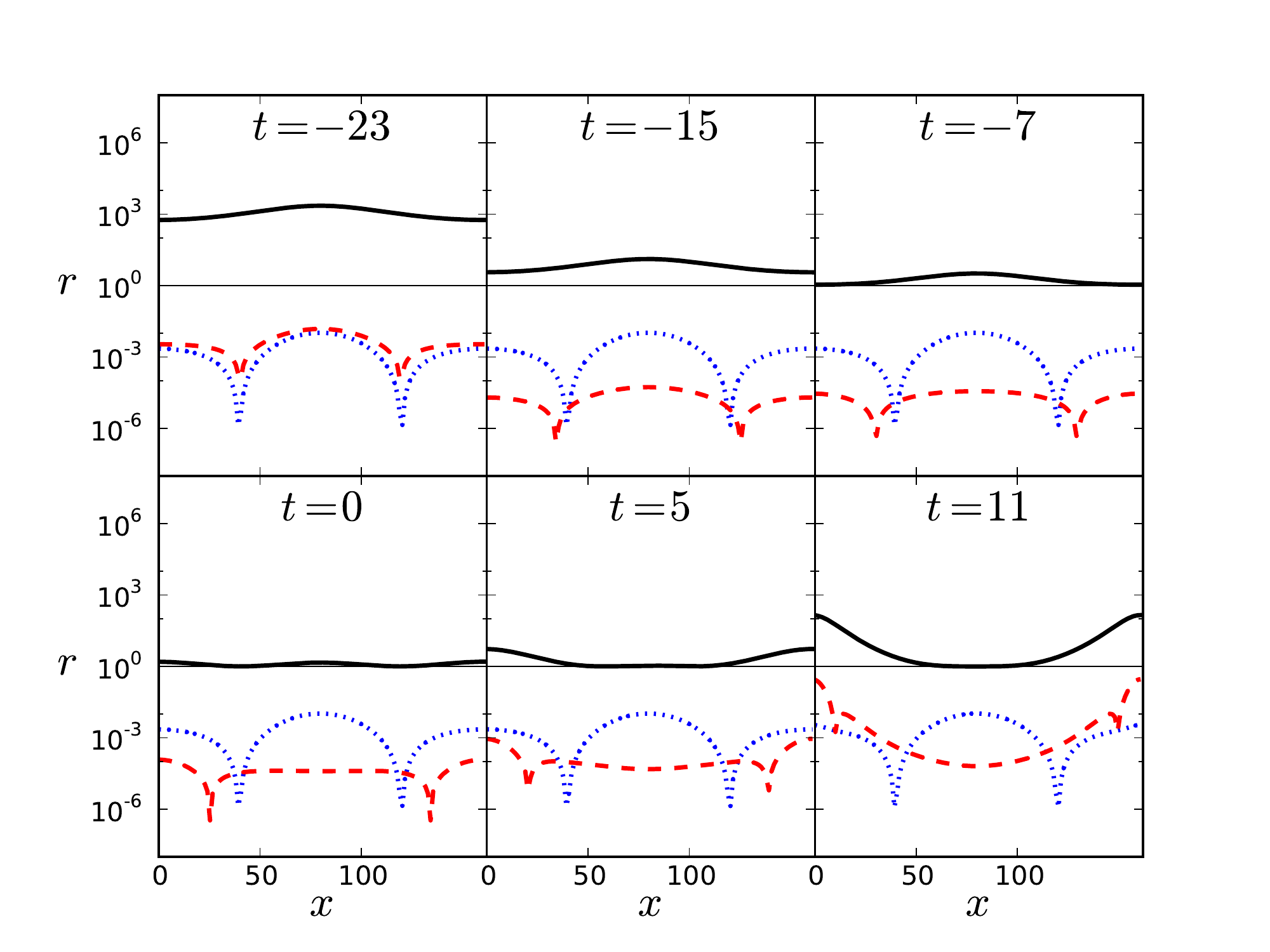}
\caption{(color online) $|E_\phi / E_\chi|$ (black continuous), $|\tfrac{1}{2} {}^{(3)}\!R / E_\chi|$ (red dashed), and $|\sigma^2 / E_\chi|$ (blue dotted) as a function of the coordinate $x$ at select times, computed with parameters in (\ref{eq:amp-2.5}). $x$ and $t$ coordinates are scaled in the same way as in Fig.~\ref{fig:theta-3}. In this marginal case, the ratio $|\sigma^2 / E_\chi|$ between the anisotropy and the $\chi$ field energy density reaches as high as $10^{-2}$, causing significant nonlinearity.} \label{fig:rho-2.5}
\end{figure}

This criterion also helps to estimate the effect of nonlinearity during the bounce. For a ratio $|\sigma^2 / E_\chi|$ less than but close to $1$, the substantial amount of anisotropy has a nonlinear effect on the bouncing process. A marginal case is given by the parameters
\begin{alignat}{2} \label{eq:amp-2.5}
f_0 &= -0.01 , \quad & f_1 &= 0.003 , \nonumber \\
f_2 &= 0.007 , \quad & f_3 &= -0.015 .
\end{alignat}
Fig.~\ref{fig:theta-2.5} shows the local expansion $\theta$ at select times,
and Fig.~\ref{fig:rho-2.5} shows the ratios $|E_\phi / E_\chi|$, $|\tfrac{1}{2} {}^{(3)}\!R / E_\chi|$, and $|\sigma^2 / E_\chi|$.
In this example, the ratio $\sigma^2 / |E_\chi|$ reaches a maximum of $\sim 10^{-2}$ near the middle of the $x$ range in the figure. Accordingly, the bounce in this region is much delayed relative to other regions, making the universe spatially inhomogeneous. The large anisotropy as compared to the $\chi$ field energy density also implies that perturbative analysis is not accurate, since in linear perturbation theory anisotropy is a second order effect that must be negligible. Therefore in this case we also expect significant nonlinear effects in the evolution of adiabatic perturbations, as discussed in Section~\ref{sec:nonlinearity}.

Our computation shows that the presence of large inhomogeneity and anisotropy with respect to the energy density of the $\chi$ field before the bouncing phase results in nonlinear growth of curvature and anisotropy that can disrupt the nonsingular bounce. On the other hand, sufficiently small perturbations can pass through the nonsingular bounce without affecting it. Nevertheless, these adiabatic modes may become strongly coupled during the bouncing phase, which can alter the power spectrum and induce non-Gaussianity. In the next section, we study the nonlinearity in the evolution of such small perturbations for which anisotropy is subdominant with respect to the scalar field energy density.

\subsection{Nonlinearity and strong coupling} \label{sec:nonlinearity}

The adiabatic perturbation is calculated using the covariant formalism, i.e., by solving for the integrated expansion $\mathcal{N}$ from Eq.~(\ref{eq:Ndot}). To quantify the magnitude of the nonlinearity, we decompose $\mathcal{N}$ into Fourier modes at each time step,
\begin{equation} \label{eq:fourier}
\mathcal{N}(t,x) = \mathcal{N}^{(0)}(t) + \mathcal{N}^{(1)}(t) \cos(mx) + \mathcal{N}^{(2)}(t) \cos(2mx) + \cdots \, .
\end{equation}
The zeroth mode gives the homogeneous part of $\mathcal{N}$, which corresponds to the background solution $N$ in Eq.~(\ref{eq:N0}). The first Fourier mode $\mathcal{N}^{(1)}$ with $k = m$ corresponds to the linear perturbation given in Eq.~(\ref{eq:N1}), which can be compared to the curvature perturbation $-\psi^\text{(h)}$ in the linear harmonic gauge, presented in Appendix~\ref{sec:linear}. The second Fourier mode $\mathcal{N}^{(2)}$ can only arise from nonlinearities in either the initial data or the evolution equations. For small perturbations, quadratic terms would be the leading nonlinear contribution in a perturbative expansion. Therefore, the amplitude of $\mathcal{N}^{(2)}$ with double wavenumber $k = 2m$ represents the leading order nonlinearity in the curvature perturbation. 

\begin{figure}
\centering
\includegraphics[width=0.5\textwidth]{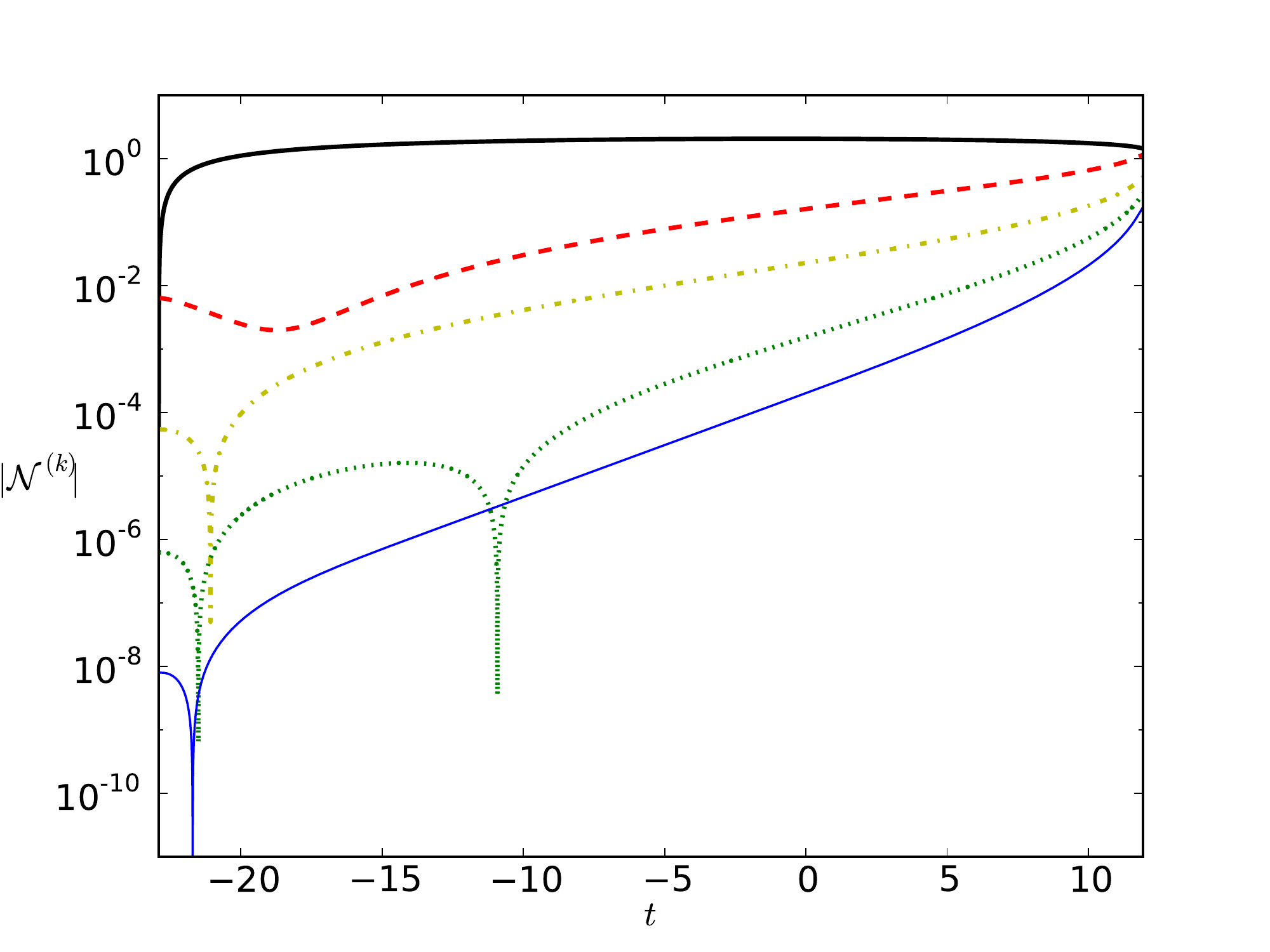}
\caption{(color online) Amplitude of the first few Fourier modes of the integrated expansion $\mathcal{N}$, computed with parameters in (\ref{eq:amp-2.5}): $|\mathcal{N}^{(0)}|$ (black thick), $|\mathcal{N}^{(1)}|$ (red dashed), $|\mathcal{N}^{(2)}|$ (yellow dash-dotted), $|\mathcal{N}^{(3)}|$ (green dotted), $|\mathcal{N}^{(4)}|$ (blue thin). $t$ is scaled in the same way as in Fig.~\ref{fig:a}, so that $t=0$ corresponds to the bounce in the homogeneous case. In this example, the amplitude of the higher Fourier modes, especially $\mathcal{N}^{(2)}$, is separated by less than 1 order of magnitude from the linear mode $\mathcal{N}^{(1)}$, indicating that nonlinearity becomes significant during the bouncing phase.} \label{fig:msn_k-nonlin}
\end{figure}

\begin{figure}
\centering
\includegraphics[width=0.5\textwidth]{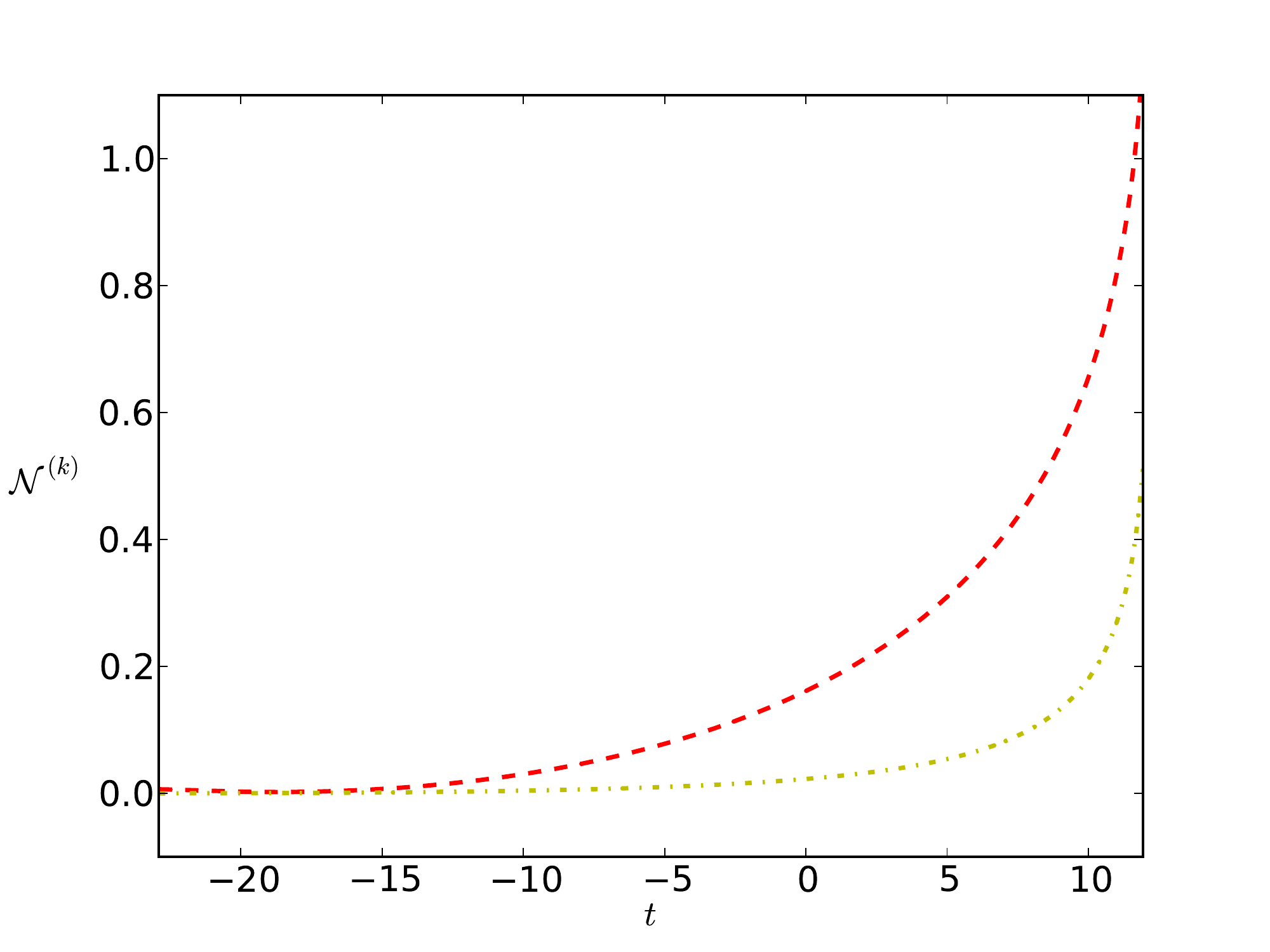}
\caption{(color online) Comparison of the Fourier modes $\mathcal{N}^{(1)}$ (red dashed) and $\mathcal{N}^{(2)}$ (yellow dash-dotted). $t$ is scaled in the same way as in Fig.~\ref{fig:msn_k-nonlin}. The amplitude of $\mathcal{N}^{(2)}$ becomes substantial during the bounce as compared to $\mathcal{N}^{(1)}$.} \label{fig:N2nonlin}
\end{figure}

Consider the previous example with parameters given in (\ref{eq:amp-2.5}). The first few Fourier modes of the integrated expansion $\mathcal{N}$ are plotted as a function of the harmonic time $t$ in Fig.~\ref{fig:msn_k-nonlin}.
It can be seen in this example that the nonlinearity is relatively large compared to the amplitude of the linear term. In particular, the amplitude of $\mathcal{N}^{(2)}$ is initially suppressed with respect to $\mathcal{N}^{(1)}$ by 2 orders of magnitude, but, after a short time, this ratio decreases to less than 1 order of magnitude, indicating that nonlinearity is no longer negligible. Similar behavior can also be observed for higher Fourier modes. Fig.~\ref{fig:N2nonlin} gives a direct comparison between the amplitude of $\mathcal{N}^{(1)}$ and $\mathcal{N}^{(2)}$,
showing that the latter rapidly grows in the bouncing phase.

\begin{figure}
\centering
\includegraphics[width=0.5\textwidth]{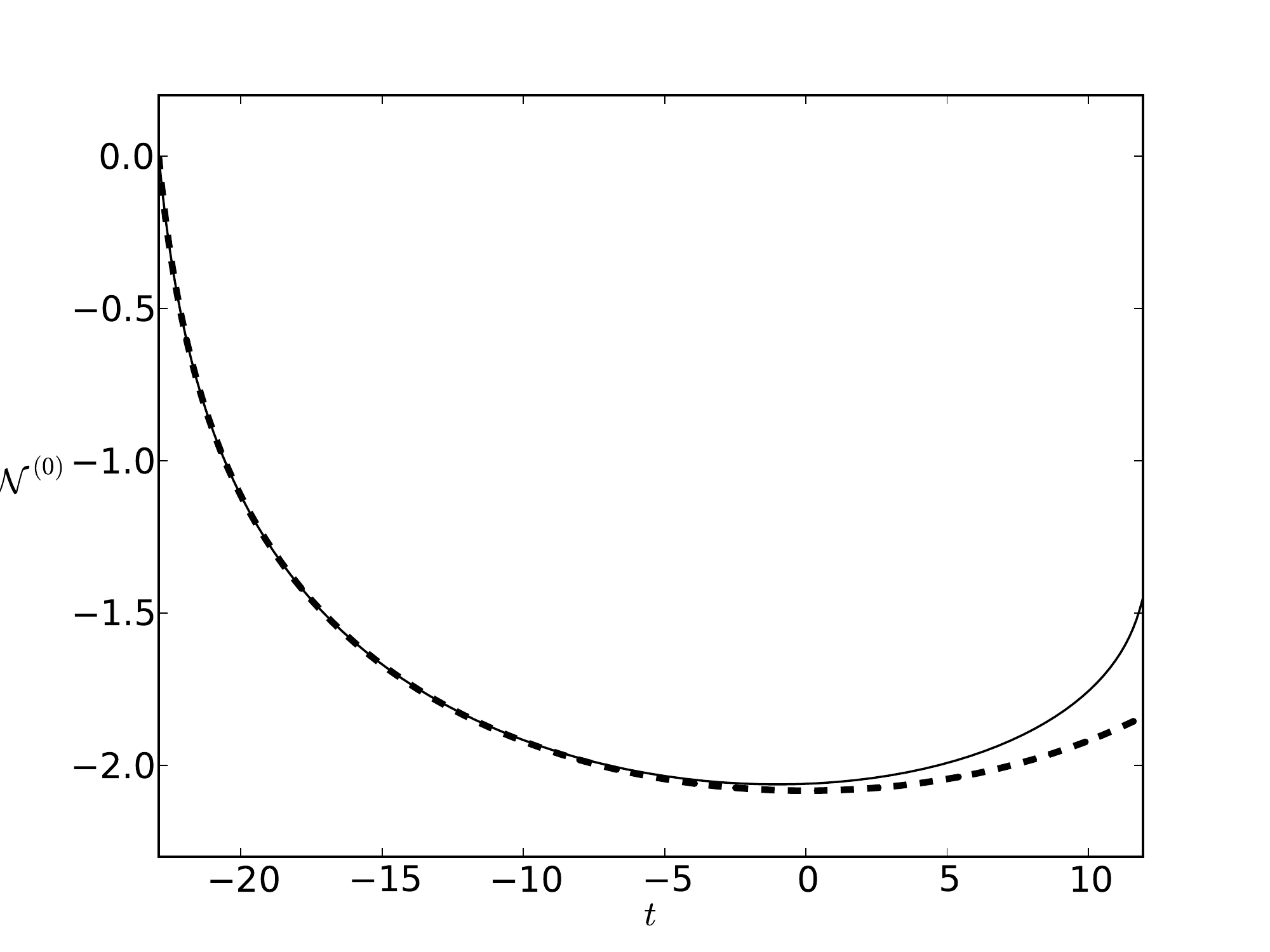}
\caption{(color online) The homogeneous part of the integrated expansion, $\mathcal{N}^{(0)}$ (continuous), as compared to the background solution $N$ (dashed). $t$ is scaled in the same way as in Fig.~\ref{fig:msn_k-nonlin}. The bouncing process in the inhomogeneous case deviates from the background solution due to the substantial anisotropy as compared to the scalar field energy density.} \label{fig:N0nonlin}
\end{figure}

\begin{figure}
\centering
\includegraphics[width=0.5\textwidth]{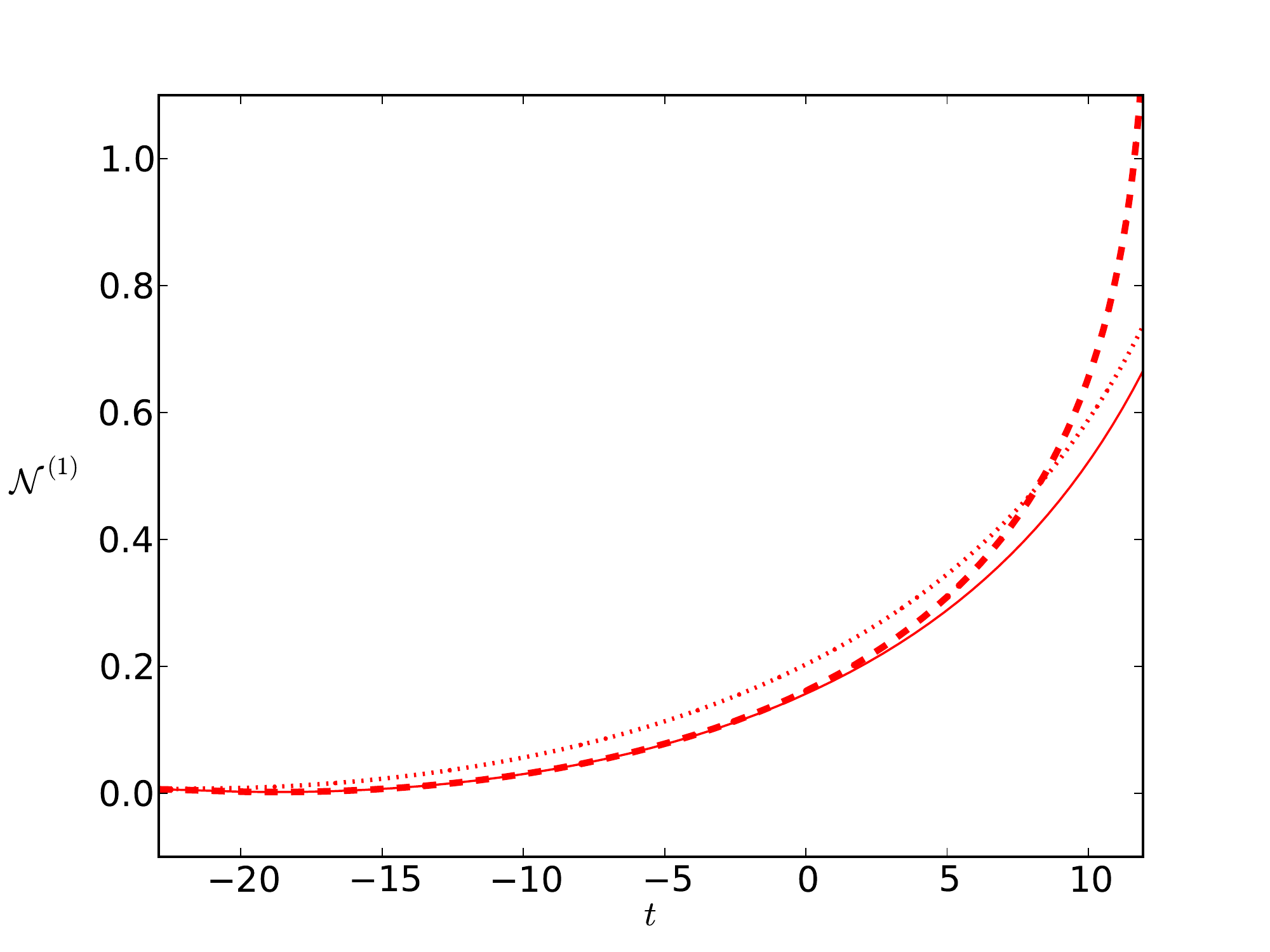}
\caption{(color online) The first Fourier mode of the integrated expansion, $\mathcal{N}^{(1)}$ (dashed), as compared to the curvature perturbation $-\psi_\text{h}$ (dotted) and lapse perturbation $\frac{1}{3} A_\text{h}$ (continuous) calculated by linear perturbation theory. $t$ is scaled in the same way as in Fig.~\ref{fig:msn_k-nonlin}. The clear disagreement between the curves indicates the inaccuracy of linear perturbative calculations due to the anisotropy.} \label{fig:N1nonlin}
\end{figure}

In addition, Fig.~\ref{fig:N0nonlin} shows the homogeneous part of $\mathcal{N}$,
which is compared to the background solution $N = \ln a$ from Section~\ref{sec:bouncing}. The clear deviation from the background solution is due to the presence of anisotropy with a considerable ratio of $|\sigma^2 / E_\chi|$ that affects the bouncing process. Fig.~\ref{fig:N1nonlin} shows the amplitude of the first Fourier mode $\mathcal{N}^{(1)}$,
as compared to the linear harmonic curvature perturbation $-\psi^\text{(h)}$ calculated in Appendix~\ref{sec:linear}. Also shown in this figure is $\frac{1}{3} A^\text{(h)}$, where $A^\text{(h)}$ is the lapse perturbation in the linear harmonic gauge, which should agree with $\mathcal{N}^{(1)}$ at linear order due to the particular gauge condition (\ref{eq:linearharm}). The disagreement between those curves indicates that calculations by linear perturbation theory are far from accurate in this case, as a result of the substantial amount of inhomogeneity and anisotropy. Note that for both perturbative and nonperturbative calculations presented in all figures, the numerical error is much smaller than the width of the curves, so the differences between the curves here represent real deviations.

In the above example, nonlinearity is mainly caused by large inhomogeneity and anisotropy as compared to the energy density of the scalar fields $\phi$ and $\chi$. On the other hand, consider sufficiently small perturbations for which anisotropy is negligible. In that case, nonlinearity would be an indicator for the strong coupling problem of the curvature perturbations. If the curvature perturbations become strongly coupled, then linear perturbation theory results would receive corrections from higher order perturbations such as $\mathcal{N}^{(2)}$. We will check the amplitude of those higher Fourier modes and assess the validity of linear perturbation theory.

\begin{figure}
\centering
\includegraphics[width=0.5\textwidth]{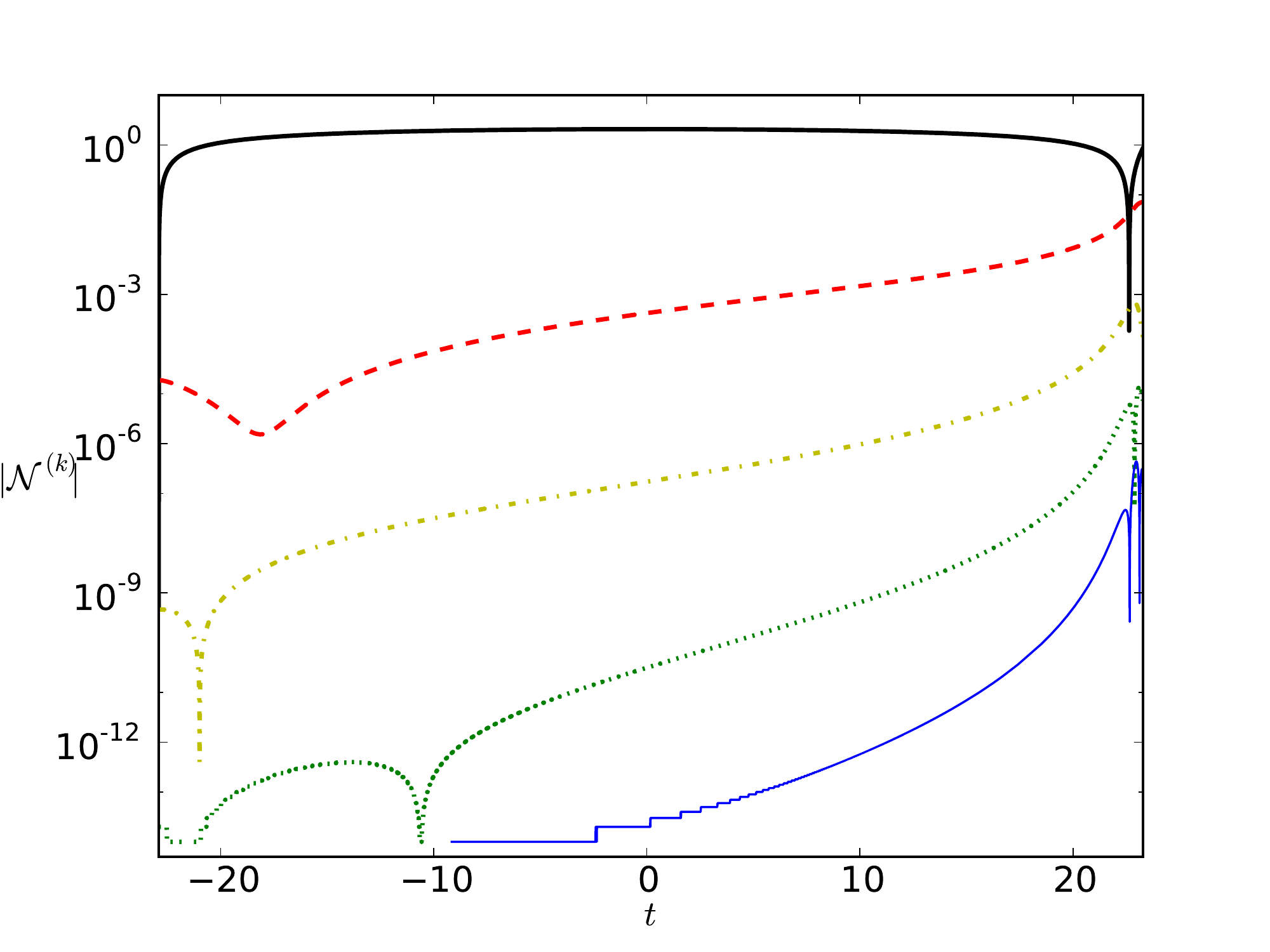}
\caption{(color online) Amplitude of the first few Fourier modes of the integrated expansion $\mathcal{N}$, computed with parameters in (\ref{eq:amp-5}): $|\mathcal{N}^{(0)}|$ (black thick), $|\mathcal{N}^{(1)}|$ (red dashed), $|\mathcal{N}^{(2)}|$ (yellow dash-dotted), $|\mathcal{N}^{(3)}|$ (green dotted), $|\mathcal{N}^{(4)}|$ (blue thin). $t$ is scaled in the same way as in Fig.~\ref{fig:a}, so that $t=0$ corresponds to the bounce in the homogeneous case. In this example, the amplitude of higher Fourier modes are clearly suppressed with respect to the linear mode, indicating that nonlinearity is negligible.} \label{fig:msn_k}
\end{figure}

\begin{figure}
\centering
\includegraphics[width=0.5\textwidth]{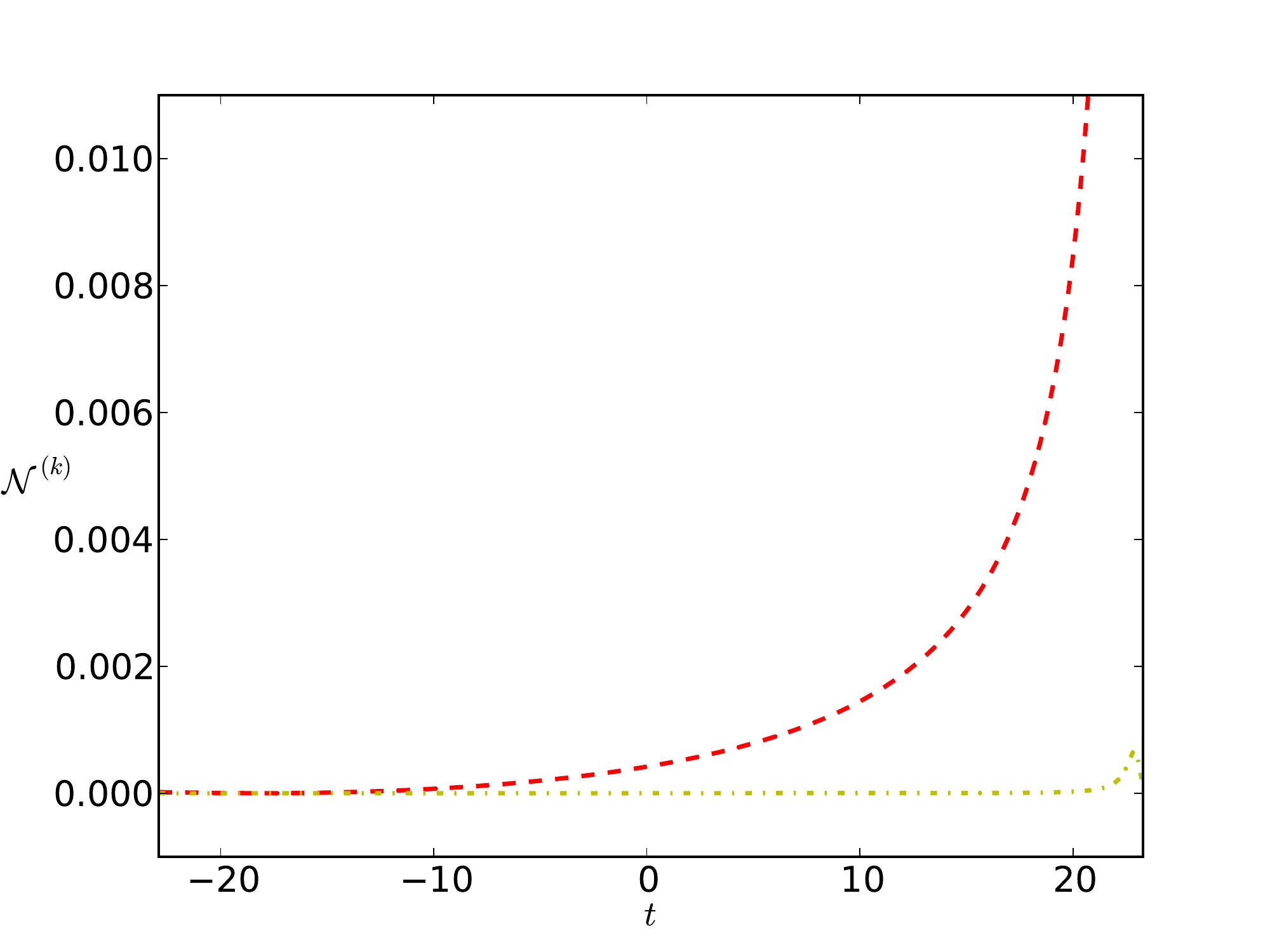}
\caption{(color online) Comparison of the Fourier modes $\mathcal{N}^{(1)}$ (red dashed) and $\mathcal{N}^{(2)}$ (yellow dash-dotted). $t$ is scaled in the same way as in Fig.~\ref{fig:msn_k}. The amplitude of $\mathcal{N}^{(2)}$ stays negligible as compared to $\mathcal{N}^{(1)}$.} \label{fig:N2}
\end{figure}

Consider an example with much smaller amplitude of perturbations compared to previous examples, with parameters
\begin{alignat}{2} \label{eq:amp-5}
m &= 0.01 , \quad & \lambda &= -0.3 , \nonumber \\
f_0 &= -0.00003 , \quad & f_1 &= 0.00001 , \\
f_2 &= 0.00002 , \quad & f_3 &= -0.00005 . \nonumber
\end{alignat}
The first few Fourier modes of the integrated expansion $\mathcal{N}$ are shown in Fig.~\ref{fig:msn_k}.
It can be seen that the higher order Fourier modes are successively suppressed by many orders of magnitude, suggesting that nonlinearity is negligible. Note that the common increase of their amplitude near the end is an artifact of the harmonic slicing -- as noted in Section~\ref{sec:harmonic}, the future infinity in physical time is compactified to a finite harmonic time, which appears to amplify the inhomogeneities. Fig.~\ref{fig:N2} compares the amplitude of the second Fourier mode $\mathcal{N}^{(2)}$ to the first Fourier mode $\mathcal{N}^{(1)}$.
The fact that $\mathcal{N}^{(2)}$ remains small compared to $\mathcal{N}^{(1)}$ throughout the bounce implies that nonlinearity is truly insignificant in this example.

\begin{figure}
\centering
\includegraphics[width=0.5\textwidth]{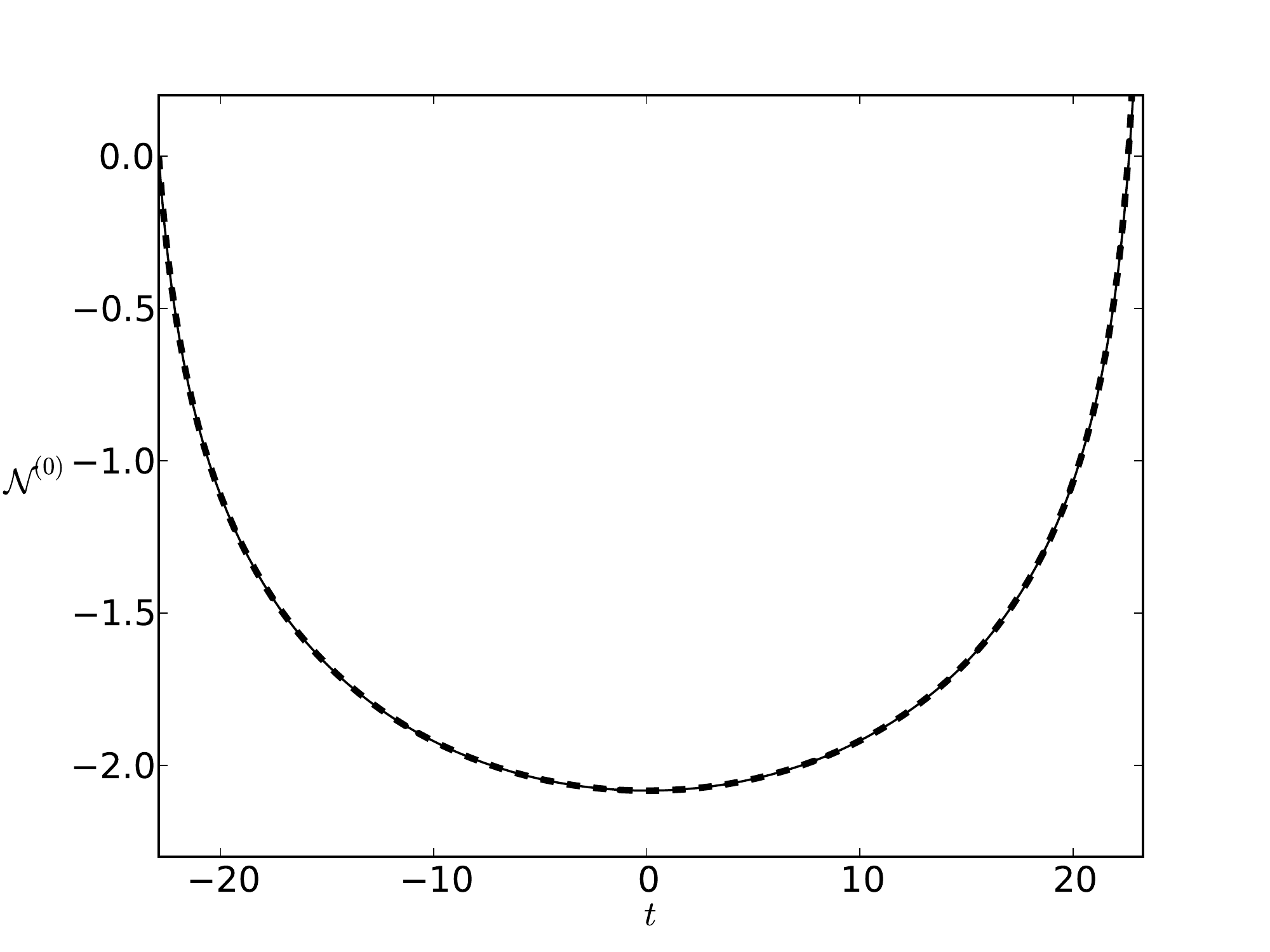}
\caption{(color online) The homogeneous part of the integrated expansion, $\mathcal{N}^{(0)}$ (continuous), as compared to the background solution $N$ (dashed). $t$ is scaled in the same way as in Fig.~\ref{fig:msn_k}. The perfect agreement shows that the background solution is a good approximation.} \label{fig:N0}
\end{figure}

\begin{figure}
\centering
\includegraphics[width=0.5\textwidth]{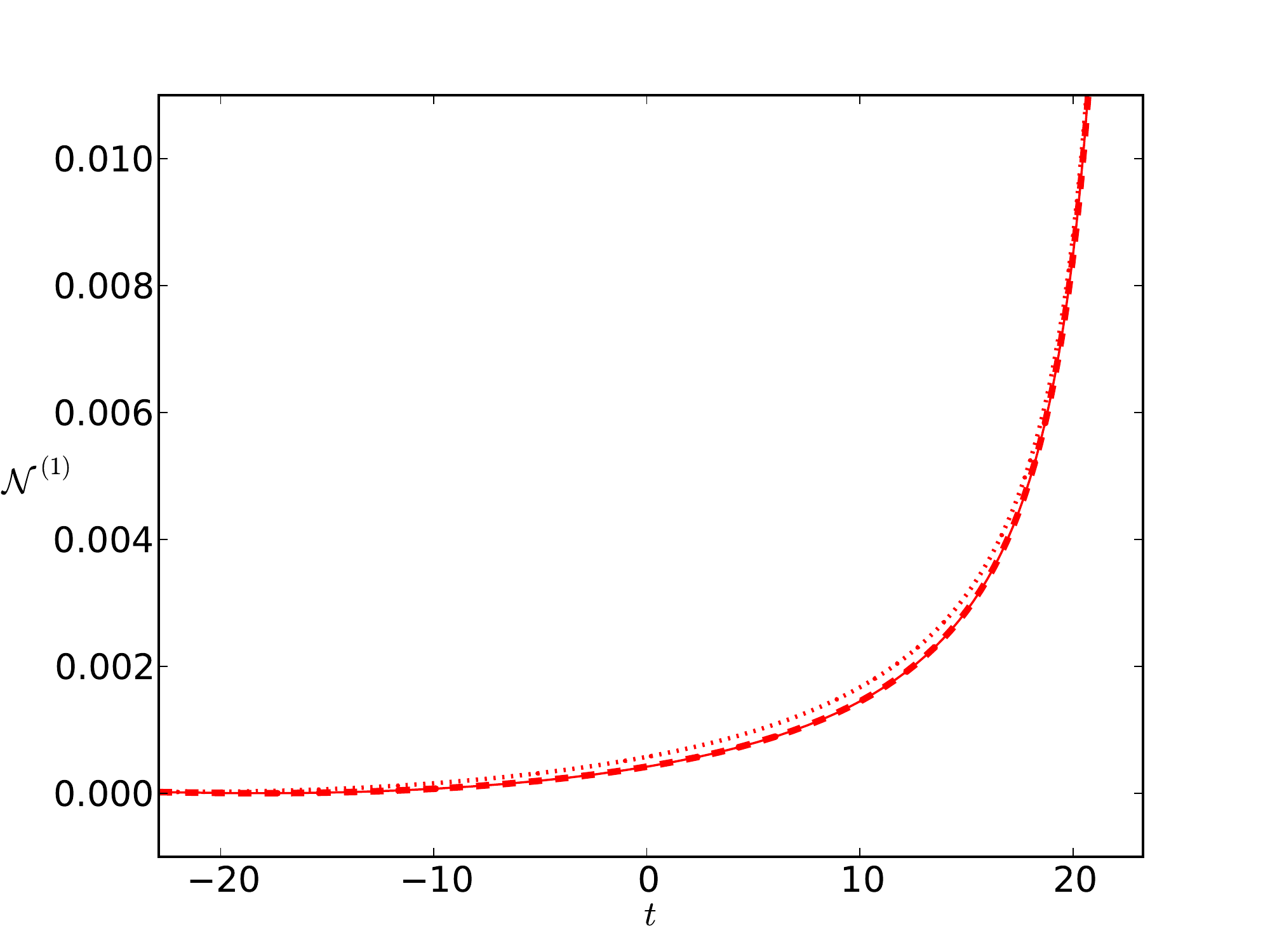}
\caption{(color online) The first Fourier mode of the integrated expansion, $\mathcal{N}^{(1)}$ (dashed), as compared to the curvature perturbation $-\psi_\text{h}$ (dotted) and lapse perturbation $\frac{1}{3} A_\text{h}$ (continuous) calculated by linear perturbation theory. $t$ is scaled in the same way as in Fig.~\ref{fig:msn_k}. All three curves agree to good approximation, showing that linear perturbation theory works well in this case.} \label{fig:N1}
\end{figure}

For this same example, the value of $\mathcal{N}^{(0)}$ is shown separately in Fig.~\ref{fig:N0},
which agrees perfectly with the background solution $N = \ln a$ from Section~\ref{sec:bouncing}. The amplitude of $\mathcal{N}^{(1)}$ is shown in Fig.~\ref{fig:N1},
together with the linear harmonic curvature perturbation $-\psi^\text{(h)}$ calculated in Appendix~\ref{sec:linear}. The small discrepancy between $\mathcal{N}^{(1)}$ and $-\psi^\text{(h)}$ is due to the gradient term in Eq.~(\ref{eq:N1}). A better agreement is shown between $\mathcal{N}^{(1)}$ and $\frac{1}{3} A^\text{(h)}$, which illustrates that linear perturbation theory gives a quite accurate result for the curvature perturbation.

The fact that the evolution of a single Fourier mode does not suffer from nonlinearity during the bounce suggests that each mode evolves independently. According to linear perturbation theory, the total curvature perturbation is a superposition of different modes. In particular, there should be no mixing between various modes. We test the linear superposition by studying the evolution of multiple modes, using the ansatz (\ref{eq:g0} - \ref{eq:A11g}). Similar to the above analysis where we follow the amplitude of double wavenumber modes to check for nonlinearity, below we focus on the amplitude of the mixed modes to address the validity of superposition.

\begin{figure}
\centering
\includegraphics[width=0.5\textwidth]{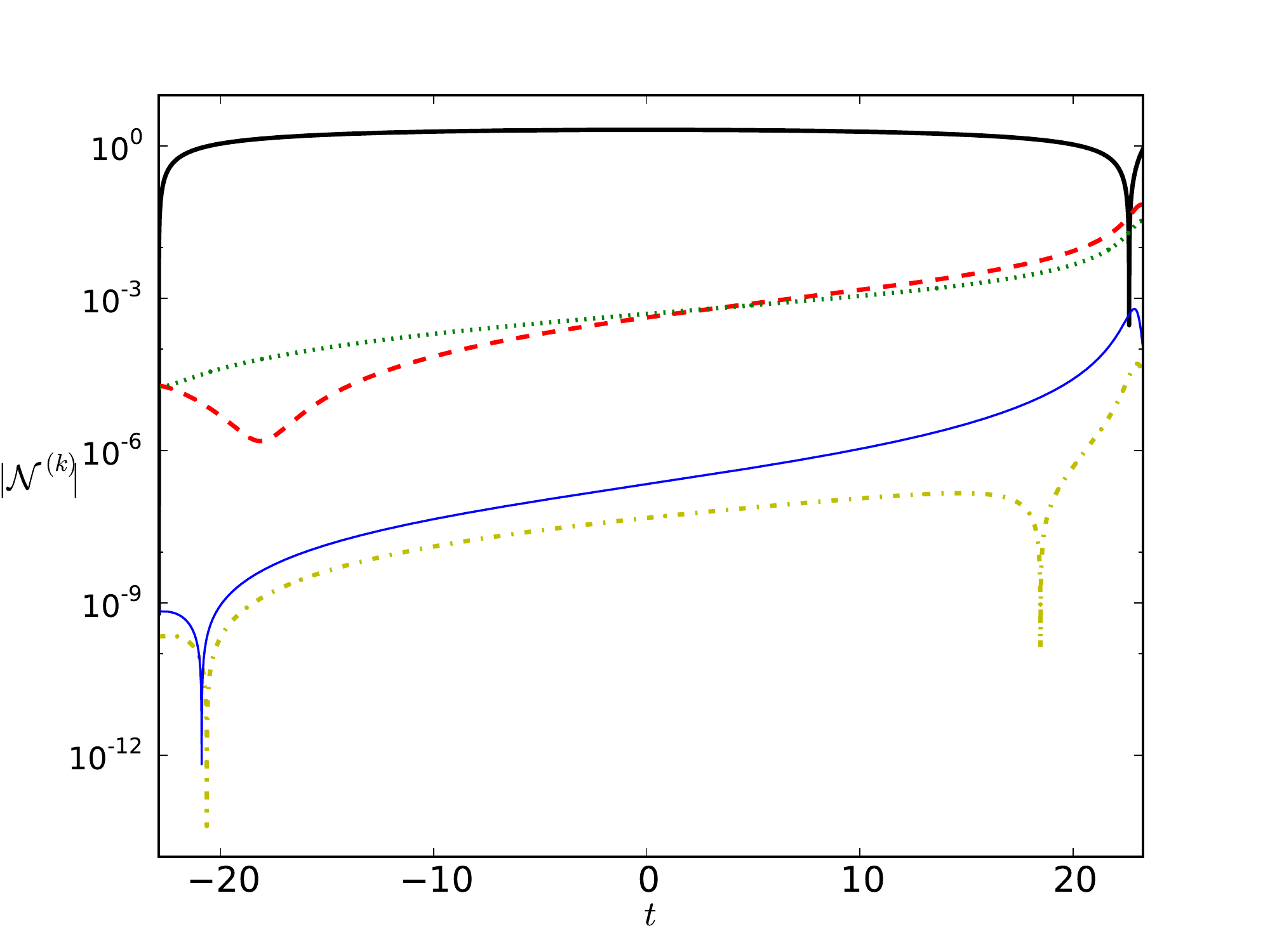}
\caption{(color online) Amplitude of the first few Fourier modes of the integrated expansion $\mathcal{N}$, computed with parameters in (\ref{eq:amp-5g}): $|\mathcal{N}^{(0)}|$ (black thick), $|\mathcal{N}^{(1)}|$ (red dashed), $|\mathcal{N}^{(2)}|$ (yellow dash-dotted), $|\mathcal{N}^{(3)}|$ (green dotted), $|\mathcal{N}^{(4)}|$ (blue thin). $t$ is scaled in the same way as in Fig.~\ref{fig:a}, so that $t=0$ corresponds to the bounce in the homogeneous case. In this example, the amplitude of the two input modes $\mathcal{N}^{(1)}$ and $\mathcal{N}^{(3)}$ are comparable to each other, whereas the mixed modes $\mathcal{N}^{(2)}$ and $\mathcal{N}^{(4)}$ are suppressed, indicating that mode mixing is negligible.} \label{fig:msn2_k}
\end{figure}

As an example, consider the parameters in (\ref{eq:amp-5}) plus
\begin{alignat}{2} \label{eq:amp-5g}
m_1 &= 0.01 . \quad & m_2 &= 0.03 . \nonumber \\
g_0 &= 0.00002 , \quad & g_1 &= -0.00001 , \\
g_2 &= -0.00002 , \quad & g_3 &= 0.00001 , \nonumber
\end{alignat}
which represent a second Fourier mode with an amplitude comparable to the mode given by (\ref{eq:amp-5}). Since $m_1 = m$ as in (\ref{eq:amp-5}) and $m_2 = 3 m$, the mixed modes have wavenumbers $k = 2 m$ and $4 m$. Following Eq.~(\ref{eq:fourier}), we decompose the integrated expansion $\mathcal{N}$ into Fourier modes with wavenumbers equal to multiples of $m$. Fig.~\ref{fig:msn2_k} shows the amplitude of the first few modes during the bounce.
The principal modes with $k = m$ and $3 m$ have comparable amplitudes, as set by the initial values, whereas the mixed modes with $k = 2 m$ and $4 m$ are clearly suppressed. This verifies that there is little mixing between different modes, consistent with the absence of nonlinearity.

\begin{figure}
\centering
\includegraphics[width=0.5\textwidth]{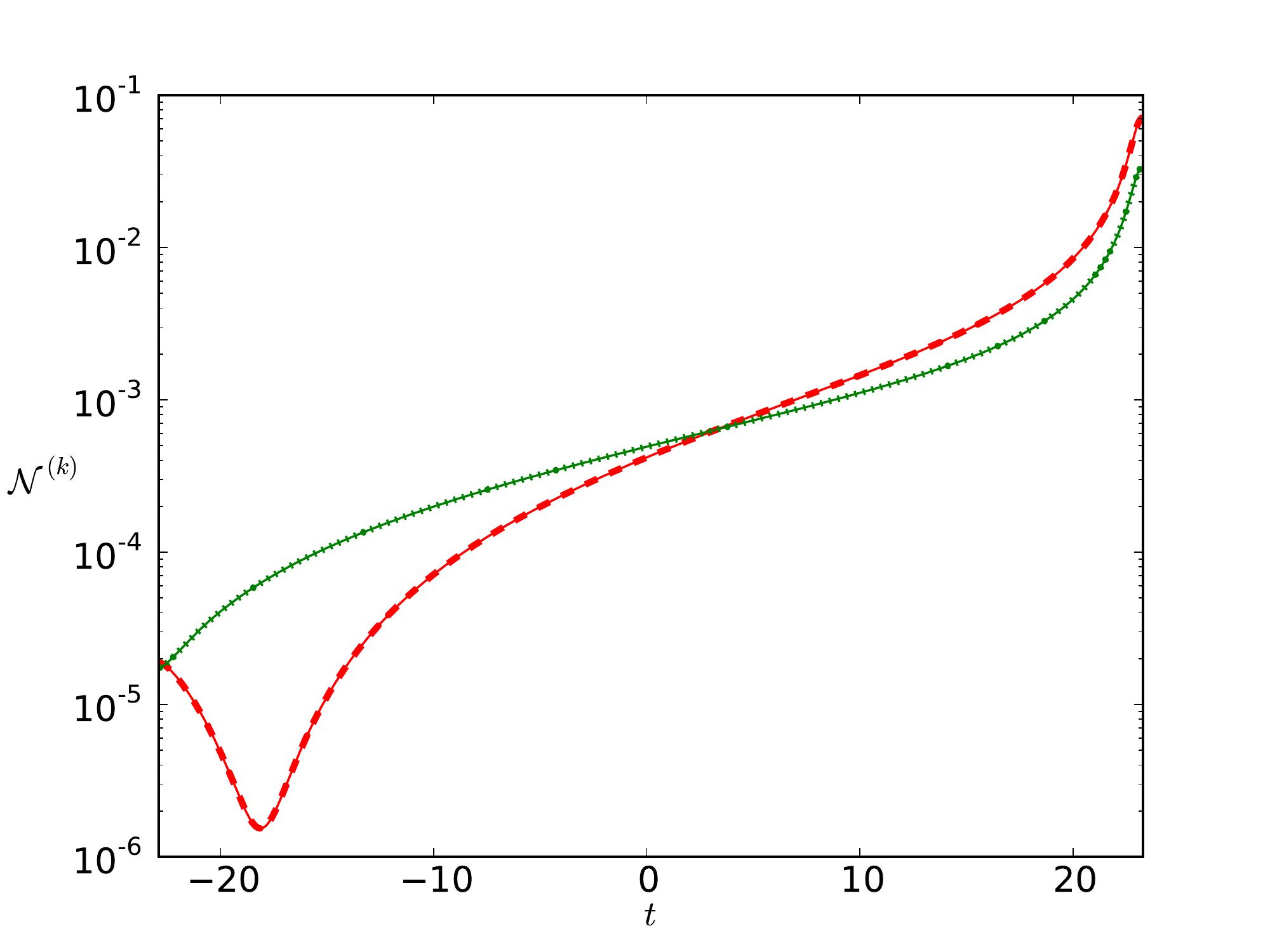}
\caption{(color online) $k = m$ mode in the two-mode computation (red dashed) and the single-mode computation (red continuous), $k = 3m$ mode in the two-mode computation (green dotted) and the single-mode computation (green continuous). $t$ is scaled in the same way as in Fig.~\ref{fig:msn2_k}. The agreement of two-mode and single-mode computations in each case shows that different Fourier modes evolve independently.} \label{fig:mode12}
\end{figure}

Moreover, we compare the evolution of each principal mode in cases with and without the presence of the other. In Fig.~\ref{fig:mode12}, the $k = m$ mode in the current example is compared to the result in the previous example where it is the only mode in the input.
Alternatively, the $k = 3 m$ mode in the current example is compared to the case with $g_i$'s given by (\ref{eq:amp-5g}) but $f_i$'s set to zero. In both comparisons the amplitude from single and double mode computations agree perfectly, confirming that different modes evolve independently regardless of one another.

Let us comment on the apparent contradiction with a naive expectation based on typical strong coupling analysis. The strong coupling argument states that, in the effective action for the curvature perturbation $\zeta$ derived from a perturbative expansion of the Einstein action, the cubic Lagrangian becomes comparable in size to the quadratic Lagrangian when the parameter $\epsilon \equiv -\dot{H} / H^2$ is much larger than $1$ \cite{Baumann:2011dt}. Such strong coupling can arise either at the classical or the quantum level, depending on whether the modes have exited the horizon. For superhorizon modes that evolve classically, strong coupling implies that the linearized equations given by the quadratic Lagrangian would receive corrections from quadratic terms given by the cubic Lagrangian; these quadratic terms become comparable to the linear terms when $\epsilon$ is large, causing the evolution to become nonlinear. In that case, solving the linearized equations would not give the correct result for the curvature perturbation and linear perturbation theory would fail. In particular, near a nonsingular bounce, $\epsilon$ diverges as $H \to 0$, implying that the quadratic terms become singular.

To understand why the argument based on strong coupling analysis fails in this case, we first note that the full equations of motion (\ref{eq:eomphi}, \ref{eq:eomchi}) and (\ref{eq:einstein}) remain regular during the bounce. Therefore a perturbative expansion in a well-defined gauge would not introduce singular terms. In fact, the anticipated singular terms would only arise in the cubic Lagrangian that is obtained by first eliminating the lapse and shift variables through the Hamiltonian and momentum constraints \cite{Maldacena:2002vr, Seery:2005wm, Chen:2006nt}. In the same way, one may choose to eliminate the lapse and shift variables in the equations of motion. For example, the momentum constraint (\ref{eq:e2}) allows one to replace the lapse $A$ with $( - \psi' + \frac{1}{2}\phi' \delta\phi - \frac{1}{2} \chi' \delta\chi ) / \mathcal{H}$, incurring a factor $1/\mathcal{H}$ that is singular at $\mathcal{H} = 0$. However, the classical equations of motion ensure that $( - \psi' + \frac{1}{2}\phi' \delta\phi - \frac{1}{2} \chi' \delta\chi )$ and $\mathcal{H}$ vanish proportionally so as to keep the lapse $A$ finite at the bounce. Hence the quadratic terms in the equations of motion involving $\psi'$ after the substitution remain much smaller than the linear terms near the bounce, despite the singular coefficient. This explains why at the classical level the nonlinearity is negligible even though the cubic Lagrangian exhibits strong coupling. (Note that the situation is different at the quantum level where the perturbations can fluctuate independently of the fixed background $\mathcal{H}$. In this paper we do not consider the quantum strong coupling as we focus on the classical evolution of adiabatic perturbations.)

\subsection{Scale dependence} \label{sec:spectrum}

Next we study the power spectrum of the adiabatic modes. Instead of computing perturbations in the harmonic gauge, we calculate the comoving curvature perturbation $\mathcal{R}$ which becomes nearly constant on superhorizon scales in the expansion phase and determines the power spectrum of primordial fluctuations. With two scalar fields, the comoving curvature perturbation $\mathcal{R}$ can be defined as \cite{Allen:2004vz}, at linear order,
\begin{equation}
\mathcal{R} \equiv \psi + \mathcal{H} \frac{\phi' \delta\phi - \chi' \delta\chi}{{\phi'}^2 - {\chi'}^2} \, ,
\end{equation}
where $\phi'$ and $\chi'$ are given by the background solution. This quantity $\mathcal{R}$ does not have a covariant generalization. However, shortly before and after the bounce, since $\chi'$ is negligible compared to $\phi'$, the value of $\mathcal{R}$ can be well approximated by the curvature perturbation $\mathcal{R}_\phi$ in the comoving $\phi$ gauge, given by Eq.~(\ref{eq:Rphi}). Therefore we can use $\mathcal{R}_\phi$ to study the power spectrum of the adiabatic perturbations.

The covariant generalization of $\mathcal{R}_\phi$ is the integrated expansion $\mathcal{N}^{(\phi)}$ on the constant $\phi$ slices. As discussed in Section~\ref{sec:covariant}, $\mathcal{N}^{(\phi)}$ cannot be calculated directly in the numerical computation. Nevertheless, for small amplitudes where anisotropy is negligible compared to the scalar field energy density, linear perturbation theory is shown to work throughout the bounce. Therefore we adopt the definition from there and use Eq.~(\ref{eq:reconstr}) to reconstruct $\mathcal{R}_\phi$. At linear order,
\begin{equation} \label{eq:reconstr1}
\mathcal{R}_\phi^{(1)} = - \mathcal{N}^{(1)} + \frac{\dot{\mathcal{N}}^{(0)}}{\dot{\phi}^{(0)}} \, \phi^{(1)} ,
\end{equation}
where the quantities on the right hand side are calculated in the harmonic gauge, and the superscript $(k)$ denotes the $k$th Fourier mode in an expansion like (\ref{eq:fourier}).

To calculate the power spectrum, the initial values for the perturbations are no longer chosen arbitrarily. Instead, the adiabatic perturbations arise from quantum fluctuations that are determined by the Bunch-Davies vacuum state when the modes are deep inside the horizon in the early contraction phase. Recall that during the contraction phase the $\chi$ field is negligible, and the $\phi$ field follows a scaling solution with a matter-like equation of state $w = 0$. For such background evolution, the scalar field perturbations \emph{in the flat gauge} are given by (see Appendix~\ref{sec:initial})
\begin{equation} \label{eq:BD}
\delta\phi_\psi \approx -i C_1(k) (t - t_{-\infty}) + C_2(k) \, ,
\end{equation}
and the same for $\delta\chi_\psi$. The Fourier coefficients $C_1(k)$ and $C_2(k)$ depend on the wavenumber $k$ as $C_1 \sim k^{-3/2}$ and $C_2 \sim k^{3/2}$. The $C_1$ term dominates at late times and represents the growing mode that carries a scale invariant power spectrum, whereas the $C_2$ term is constant and subdominant. The value for $C_2$ is related to $C_1$ by Eq.~(\ref{eq:C2overC1}),
\begin{equation} \label{eq:C2}
C_2(k) = \frac{8}{3} \bigg| \frac{k}{a H} \bigg|^3 \bigg( \frac{-2}{3 \mathcal{H}} \bigg) C_1(k) \, ,
\end{equation}
and the value for $C_1$ should be normalized such that the final amplitude of the adiabatic perturbation matches the observed power spectrum of the primordial fluctuations (see Eq.~(\ref{eq:power-spectrum})). For now, we take
\begin{equation} \label{eq:C1}
C_1(k) \sim \frac{3.5 \times 10^{-5}}{(k L)^{3/2}} \, ,
\end{equation}
and bear in mind that it should be rescaled to the proper value in the end.

Thus, for our calculation that starts in the late contraction phase, the initial values for the scalar field perturbations in the flat gauge are
\begin{align}
\delta\phi_\psi(0) &= \delta\chi_\psi(0) = -i C_1(k) \Big( \frac{-2}{3\mathcal{H}_0} \Big) + C_2(k) \, , \\[4pt]
\delta\phi_\psi'(0) &= \delta\chi_\psi'(0) = -i C_1(k) \, ,
\end{align}
The initial values for $\delta\phi$, $\delta\phi'$, $\delta\chi$, and $\delta\chi'$ in the harmonic gauge are obtained through a linear transformation
\begin{align}
\delta\phi_\psi(0) &= \delta\phi(0) + \frac{\phi_0'}{\mathcal{H}_0} \psi(0) \, , \\[4pt]
\delta\phi_\psi'(0) &= \delta\phi'(0) - \bigg( \frac{a_0^6 V_0 \phi_0' - c a_0^6 V_0 \mathcal{H}_0}{\mathcal{H}_0^2} \bigg) \psi(0) + \frac{\phi_0'}{2 \mathcal{H}_0} \big( \phi_0' \delta\phi(0) - \chi_0' \delta\chi(0) \big) \, , \\[4pt]
\delta\chi_\psi(0) &= \delta\chi(0) + \frac{\chi_0'}{\mathcal{H}_0} \psi(0) \, , \\[4pt]
\delta\chi_\psi'(0) &= \delta\chi'(0) - \bigg( \frac{a_0^6 V_0 \chi_0'}{\mathcal{H}_0^2} \bigg) \psi(0) + \frac{\chi_0'}{2 \mathcal{H}_0} \big( \phi_0' \delta\phi(0) - \chi_0' \delta\chi(0) \big) \, ,
\end{align}
where $\psi(0)$ is given in terms of $\delta\phi'(0)$, $\delta\chi'(0)$, and $\delta\phi(0)$ by Eq.~(\ref{eq:CMClin}),
\begin{equation}
\psi(0) = - \frac{1}{2 a_0^4 k^2} \big( \phi_0' \delta\phi'(0) - \chi_0' \delta\chi'(0) - c a_0^6 V_0 \delta\phi(0) \big) \, .
\end{equation}

To account for the two modes $C_1$ and $C_2$, we use the ansatz (\ref{eq:g0} - \ref{eq:A11g}) with the same wavenumber $m_1 = m_2 = m$, but different phases $d_1 = - \frac{\pi}{2}$ and $d_2 = 0$. The $C_2$ mode corresponds to the amplitude of $\cos(mx)$ in a Fourier expansion as before, whereas the $C_1$ mode corresponds to the amplitude of $\sin(mx)$ (see Eq.~(\ref{eq:cos+sin})). The parameters $f_0$ - $f_3$ are inferred from the $C_1$ term in $\delta\phi(0)$, $\delta\phi'(0)$, $\delta\chi(0)$, and $\delta\chi'(0)$ by using Eqs.~(\ref{eq:linearAmp3}, \ref{eq:linearAmp4}) and (\ref{eq:linearPsi}) in Appendix~\ref{sec:linear}, and similarly for $g_0$ - $g_3$ from the $C_2$ term. Note that the last term in Eq.~(\ref{eq:A11g}) vanishes because the Bunch-Davies initial values ensure that the coefficient $f_0 g_1 - f_1 g_0 - f_2 g_3 + f_3 g_2$ equals $0$ identically.

\begin{figure}
\centering
\includegraphics[width=0.5\textwidth]{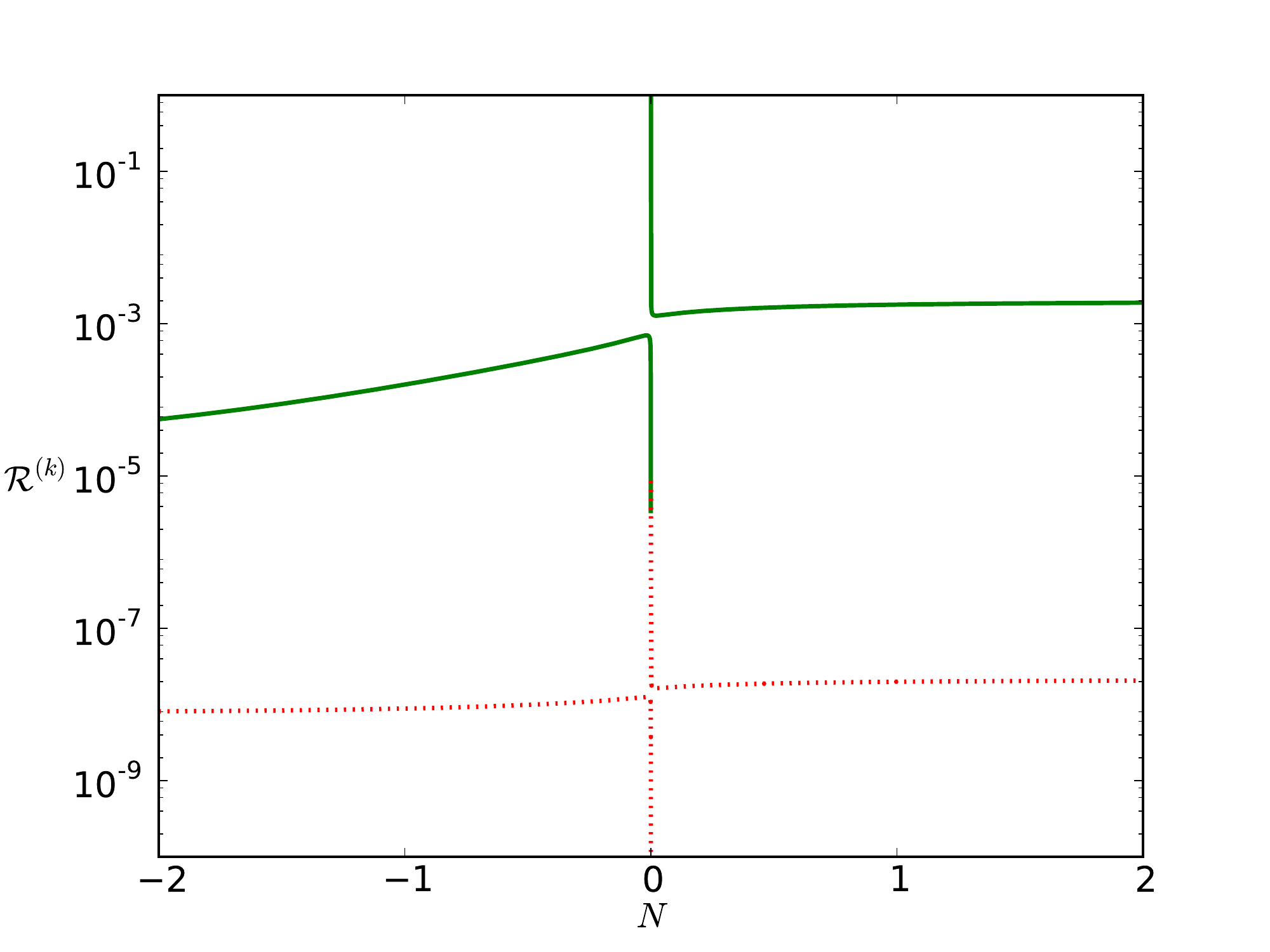}
\caption{(color online) Amplitude of the growing mode (green continuous) and the constant mode (red dotted) with Bunch-Davies initial values for $m = 0.01$. The growing mode gives the dominant contribution to the amplitude after the bounce.} \label{fig:C1C2}
\end{figure}

As an example, we calculate the growing ($C_1$) mode and the constant ($C_2$) mode for $m = 0.01$. The comoving curvature perturbation $\mathcal{R}_\phi$ for each mode is extracted from the integrated expansion $\mathcal{N}$ and the scalar field $\phi$ according to Eq.~(\ref{eq:reconstr1}). Their amplitudes are shown in Fig.~\ref{fig:C1C2} as a function of the number of e-folds $N$,
which is shifted such that the bounce occurs at $N = 0$, and the sign is chosen so that $N$ is negative before and positive after the bounce. Note that the spikes near the bounce are due to the moment when $\dot{\phi}^{(0)} = 0$ in Eq.~(\ref{eq:reconstr1}), which again illustrates that $\mathcal{N}^{(\phi)}$ cannot be evolved directly through the bounce.

The result shows that the growing mode in the contraction phase becomes constant quickly after the bounce, and remains dominant in the expansion phase. Meanwhile, the constant mode in the contraction phase also contributes a constant amplitude in the expansion phase, but is negligible compared to the contribution from the growing mode. This matching condition implies that the power spectrum of primordial fluctuations in the expansion phase is primarily determined by the growing mode in the contraction phase, in agreement with \cite{Finelli:2001sr, Allen:2004vz}. Note that the perturbation amplitude grows even after horizon crossing because the matter-like contraction phase is \emph{not} an attractor. In addition, the asymmetry of the evolution before and after the bounce is due to the entropic perturbations between the two scalar fields. Such entropic perturbations source the adiabatic perturbations near the bounce when the $\chi$ field is significant. After the bounce, the $\chi$ field energy quickly diminishes, and the adiabatic perturbation approaches a constant.

To check the scale invariance of the power spectrum, we analyze the dependence of the perturbation amplitude on the wavenumber $k$. In the matter-like contraction phase after horizon crossing, the growing mode ($C_1$) has the correct $k$-dependence, $C_1 \sim k^{-3/2}$. To maintain the scale invariance into the expansion phase, modes with different wavenumbers must have the same factor of amplification during the bounce. Otherwise, a slight change in the scale dependence would induce a small tilt or running in the power spectrum.

\begin{figure}
\centering
\includegraphics[width=0.5\textwidth]{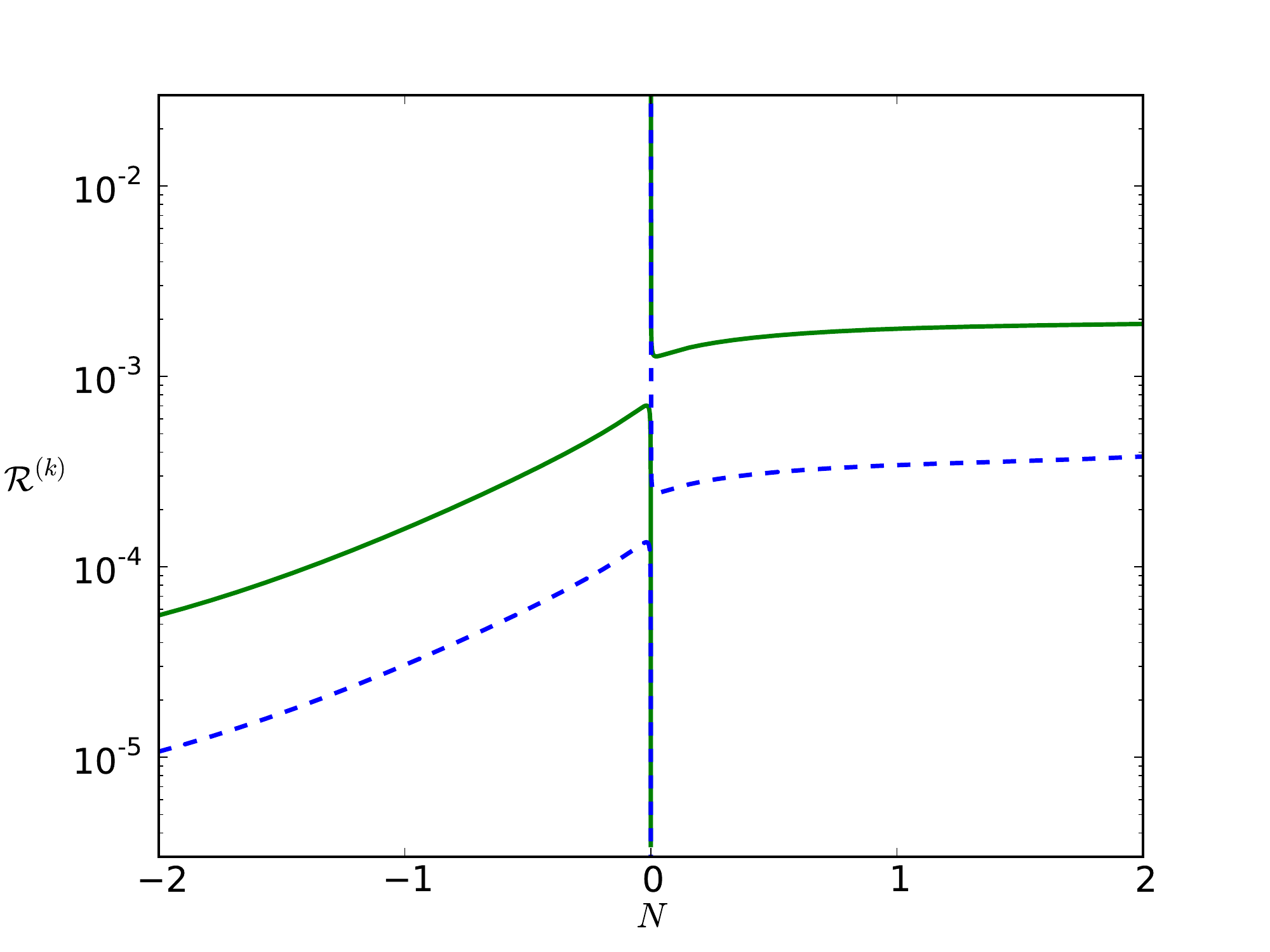}
\caption{(color online) Amplitude of the comoving curvature perturbation $\mathcal{R}_\phi$ for wavenumbers $m_1 = 0.01$ (green continuous) and $m_2 = 0.03$ (blue dashed). Their relative amplitude, given by the vertical distance, stays approximately the same after the bounce, showing that the scale dependence of the evolution of the perturbation amplitudes during the bounce is weak.} \label{fig:m1m2}
\end{figure}

\begin{figure}
\centering
\includegraphics[width=0.5\textwidth]{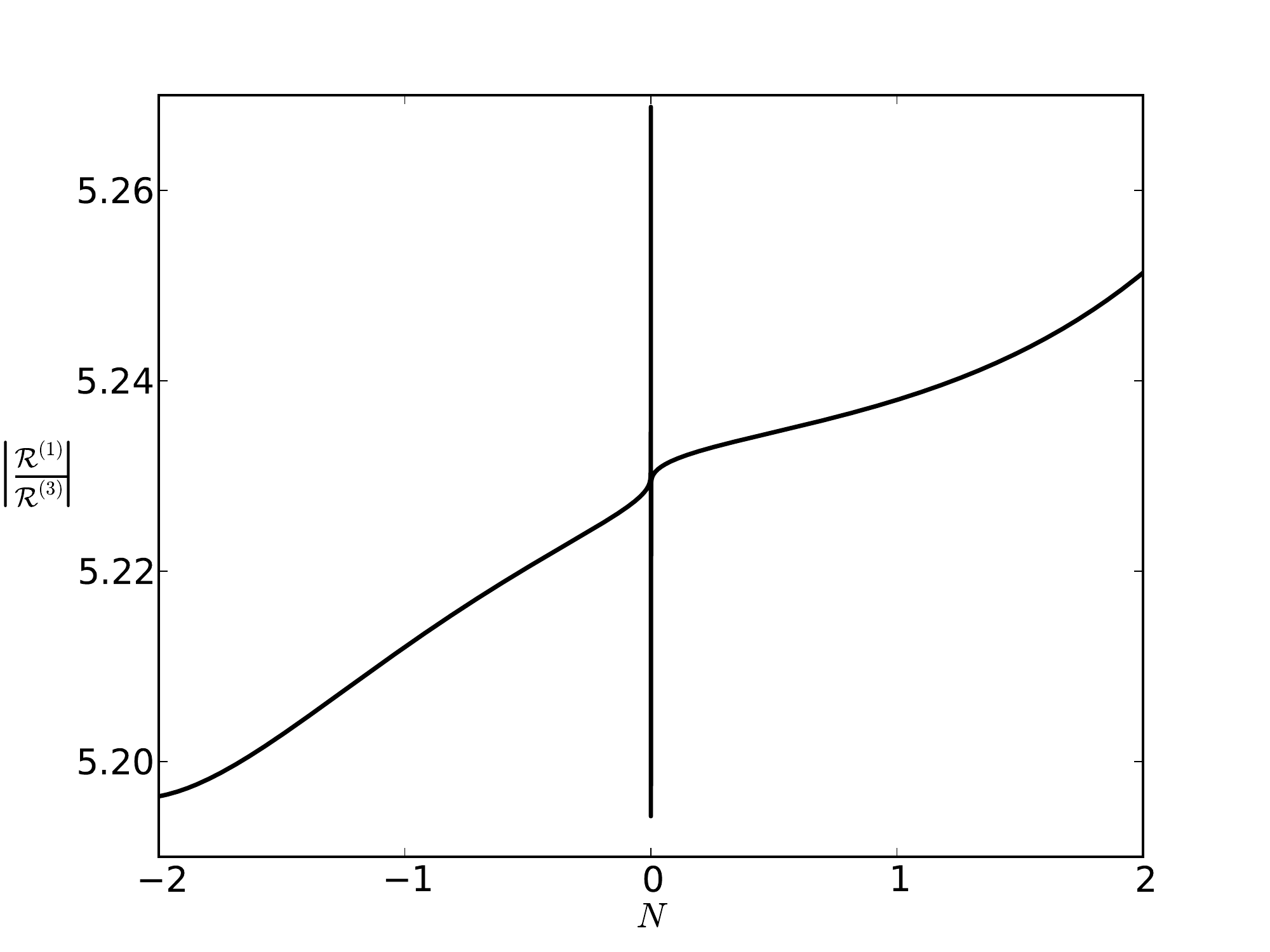}
\caption{The ratio between the amplitude of the Fourier modes with $k = m$ and $k = 3 m$. The increase of the ratio after the bounce implies a small red tilt (though negligible for observable modes, see Fig.~\ref{fig:spectrum}).} \label{fig:amp_ratio}
\end{figure}

As a case study, we calculate the comoving curvature perturbation $\mathcal{R}_\phi$ for wavenumbers $k = m$ and $3 m$, where $m = 0.01$. Since the amplitude is dominated by the growing mode for each $k$, in the ansatz (\ref{eq:g0} - \ref{eq:A11g}) we include only the $C_1$ term for each of the two wavenumbers $m_1 = 0.01$ and $m_2 = 0.03$, with $d_1 = d_2 = - \frac{\pi}{2}$. Fig.~\ref{fig:m1m2} shows their amplitude on a logarithmic scale;
their relative amplitude, given by the vertical distance between the curves, stays approximately the same before and after the bounce. This suggests that the amplification factor for the perturbation amplitude depends very weakly on the wavenumber $k$. Hence, the power spectrum remains nearly scale invariant after the bounce. Fig.~\ref{fig:amp_ratio} shows the ratio between the amplitude of the two modes.
The slight increase of this ratio after the bounce indicates a slightly larger amplitude at long wavelengths, hence a red tilt. However, as shown below, the amount of tilt turns out to be negligible on observable scales.

\begin{figure}
\centering
\includegraphics[width=0.5\textwidth]{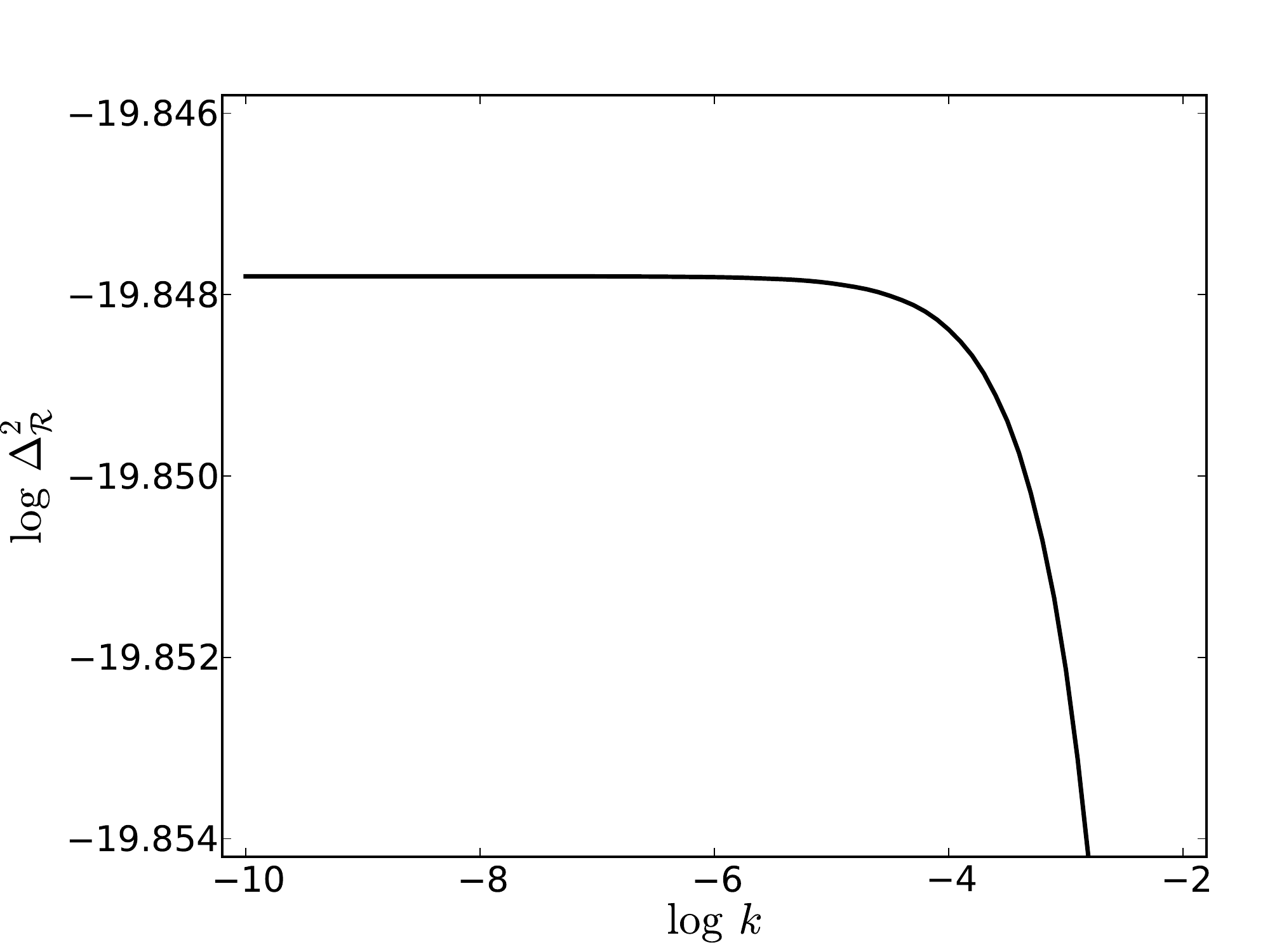}
\caption{The power spectrum of the comoving curvature perturbation $\mathcal{R}_\phi$ after the bounce. The change in the amplitude over $\sim 10$ e-folds of wavenumbers is as small as $10^{-3}$, and becomes negligible for even smaller $k$ that corresponds to observable modes.} \label{fig:spectrum}
\end{figure}

In order to quantify the deviation from scale invariance, we calculate the amplitude of the comoving curvature perturbation for a number of $k$ modes ranging over $\sim 10$ e-folds. Since their evolution is well in the linear regime, we simply use the perturbative calculation presented in Appendices~\ref{sec:linear} and \ref{sec:initial}. The power spectrum is given by Eq.~(\ref{eq:power-spectrum}), $\Delta_\mathcal{R}^2 \sim k^3 |\mathcal{R}|^2$, as shown in Fig.~\ref{fig:spectrum}.
The amplitude has been rescaled to match the observed value $\Delta_\mathcal{R}^2 \approx 2.4 \times 10^{-9}$ in the limit $k \to 0$. The absolute value of the numbers on the $\log k$ axis represents the number of e-folds after the mode exits the horizon and before the bouncing phase. The deviation of the power spectrum from a straight line indicates that there is a running of the spectral index on top of a tilt. However, the change in the amplitude is as small as $\sim 10^{-3}$ over $\sim 10$ e-folds that are shown in the figure. Such changes become negligible for even smaller $k$, especially the modes with $|\log k| \sim 50$ - $60$ that are measured in the CMB. Therefore, in practice, there is no observable spectral tilt or running.

The conservation of the power spectrum across the bounce can be understood from the smooth evolution of the long wavelength modes through the bounce. If the long wavelength modes stay outside the ``horizon'' during the bounce, then their dynamics barely depends on their wavenumbers. Here the ``horizon'' scale is not simply $1/aH$, which becomes infinite at a nonsingular bounce. Instead, it should represent the length scale at which the spatial gradient terms in the equations of motion become negligible compared to the time derivatives. A good estimate may come from the evolution equation for linear perturbations, e.g., Eq.~(\ref{eq:mukhanov}) in Appendix~\ref{sec:initial}, in which $k^2$ should be compared to $a''/a = a^2 (\dot{H} + 2 H^2)$. This last quantity does not vanish near the bounce since $\dot{H}$ is positive during the bouncing phase. It can be checked that the wavenumbers $k$ considered in our computations are indeed much smaller than $a (\dot{H} + 2 H^2)^{1/2}$ throughout the bounce.

\section{Conclusion} \label{sec:conclusion}

We have presented the first nonperturbative calculation that tracks cosmic evolution through a non-singular bounce. Our computation is based on a bouncing model with one canonical scalar field that drives a matter-like ($w=0$) contraction phase and another ghost field that induces a nonsingular bounce. We have shown that large inhomogeneity and anisotropy compared to the energy density of the ghost field can disrupt the bounce. Nonlinear effects become substantial when the anisotropy is close to or larger than the ghost field energy density that is responsible for inducing the bounce. For smaller perturbations, the anisotropy remains subdominant and does not affect the nonsingular bounce. In those cases, such as one with an amplitude consistent with observed primordial fluctuations, nonlinearities are insignificant during the bounce and the strong coupling problem does not occur for superhorizon modes, indicating that the nonsingular bounce does not cause large non-Gaussianity. We have further analyzed the scale dependence of the amplitude of the adiabatic perturbations and showed that, given scale invariant amplitudes generated in the matter-like contraction phase, the power spectrum remains scale invariant in the expansion phase without observable deviations, consistent with current observational constraints.

A new picture that emerges from our study is that the nonsingular bounce can happen in separate parts of the universe. Specifically, regions of the universe that are overwhelmed by inhomogeneity and anisotropy collapse into singularities, whereas regions with relatively smooth and isotropic conditions pass through a nonsingular bounce. This gives a completely different global picture of a nonsingular bouncing universe from what has been expected by linear perturbative analysis. The new scenario resembles the ``phoenix universe'' model \cite{Lehners:2009eg} in which a contracting universe collapses in certain regions and bounces in the others, except that here the bounce is nonsingular. Since the inhomogeneous regions of the universe terminate in singularities (barring quantum effects), the volume of the universe is dominated by the desirable regions that pass through the bounce and expand. One can further imagine that the local amplitude of the primordial fluctuations within an expanding region that bounced successfully evolves indifferently to the existence of collapsed regions that are way beyond the horizon. Since the evolution of different regions of the universe can be followed through the bounce by directly solving classical equations of motion, it may be possible to find a definite probability measure for various observables over all bouncing regions.

There are some drawbacks in the specific model considered in this paper, especially that the matter-like contraction is not an attractor solution. Consequently, a sufficiently long period of matter-like contraction phase requires fine-tuning of initial values for the background solution. Moreover, classical inhomogeneity and anisotropy grow faster than the background energy density with a matter-like equation of state, so it requires further fine-tuning to suppress inhomogeneities during the contraction phase. Furthermore, although the matter-like contraction phase can create scale invariant adiabatic perturbations, the dominant mode of these perturbations that carries the scale invariant power spectrum is not conserved even on superhorizon scales; hence the amplitude of the adiabatic perturbations in the expansion phase is different from that at the horizon crossing.

The problem with the growth of inhomogeneity and anisotropy can be avoided in an ekpyrotic contraction phase. The ekpyrotic contraction is an attractor that automatically smooths away initial inhomogeneity, spatial curvature, and anisotropy \cite{Khoury:2001wf, Erickson:2003zm, Garfinkle:2008ei}. It may be possible to have such an ekpyrotic phase before (or after, e.g., in \cite{Cai:2012va}) the matter-like contraction phase to provide extremely homogeneous and isotropic conditions for the latter. Nevertheless, the initial values for the scalar fields would still require fine-tuning in order for the homogeneous solution to stay close to the scaling solution during the entire matter-like contraction phase; the non-conservation of the growing mode of the adiabatic perturbations outside the horizon still exists as well.

Alternatively, it is possible to completely replace the matter-like contraction phase by an ekpyrotic contraction phase. Instead of using the matter-like contraction phase to generate scale invariant adiabatic perturbations, the latter can be generated through an entropic mechanism with the existing two scalar fields \cite{Gordon:2000hv, Koyama:2007mg, Lehners:2007ac, Buchbinder:2007ad, Creminelli:2007aq}. The entropic perturbations between the two fields can be converted into adiabatic perturbations at the end of the ekpyrotic phase, which then become conserved outside the horizon. Based on our results, we expect the adiabatic perturbations to evolve through the nonsingular bounce without altering the scale invariance of the power spectrum. This new scenario will be pursued elsewhere \cite{Xue:thesis}.

The remaining issue is the quantum instability of the ghost field that is used to violate the NEC and induce the nonsingular bounce. The ghost field serves as an effective mechanism for describing the bouncing process and studying the classical perturbations on superhorizon scales. To complete the model, the ghost field must be stabilized by some unknown UV-completion mechanism which is not considered here. The positive results of our above analysis inspire us to look for more realistic mechanisms of creating a nonsingular bounce that is free from the ghost instability.

\begin{acknowledgments}
BKX thanks David Wands for helpful discussions on the nonsingular bouncing model used in this paper. DG is supported by NSF grants No. PHY-0855532 and PHY-1205202. FP is supported by NSF Grants No. PHY-0745779, PHY-1065710, PHY-1305682, and the Simons Foundation. PJS is supported by the US Department of Energy Grant DE-FG02-91ER40671.
\end{acknowledgments}

\appendix

\section{Linear perturbations in harmonic gauge} \label{sec:linear}

Here we present the calculation of linear perturbations in the harmonic gauge. The background solution in harmonic time $t$ is given in Section~\ref{sec:bouncing}. Consider the following metric with scalar perturbations,
\begin{equation}
ds^2 = - a^6 (1 + 2 A) dt^2 + 2 a^4 B_{,i} dt dx^i + a^2 \big( (1 - 2 \psi) \delta_{ij} + 2 E_{,ij} \big) dx^i dx^j .
\end{equation}
After computing the Christoffel symbols, one finds, to linear order,
\begin{align}
g^{\mu\nu} \Gamma^0_{\mu\nu} &= - \frac{1}{a^6} \big[ A' + 3 \psi' - \nabla^2 (E' - a^2 B) \big] \, , \\[4pt]
g^{\mu\nu} \Gamma^i_{\mu\nu} &= - \frac{1}{a^6} \big[ (a^2 B)' + a^4 (A - \psi - \nabla^2 E) \big]^{,i} \, .
\end{align}
Therefore the harmonic gauge condition (\ref{eq:gGamma}) is specified by the constraints
\begin{align}
\mathcal{C}^0 &\equiv A' + 3 \psi' - \nabla^2 (E' - a^2 B) = 0 \, , \label{eq:linearharm} \\[4pt]
\mathcal{C} &\equiv (a^2 B)' + a^4 (A - \psi - \nabla^2 E) = 0 \, . \label{eq:linearharmspatial}
\end{align}

Under an infinitesimal coordinate transformation
\begin{equation}
t \to t + \xi^0 , \quad x^i \to x^i + \xi^{,i} \, ,
\end{equation}
the metric perturbations become
\begin{align}
A &\to A - 3 \mathcal{H} \xi^0 - (\xi^0)' \, , \\[4pt]
B &\to B + a^2 \xi^0 - \tfrac{1}{a^2} \xi' \, , \\[4pt]
\psi &\to \psi + \mathcal{H} \xi^0 \, , \\[4pt]
E &\to E - \xi \, ,
\end{align}
hence the constraints $\mathcal{C}^0, \mathcal{C}$ become
\begin{align}
\mathcal{C}^0 &\to \mathcal{C}^0 - (\xi^0)'' + a^4 \nabla^2 \xi^0 \, , \\[4pt]
\mathcal{C} &\to \mathcal{C} - \xi'' + a^4 \nabla^2 \xi \, .
\end{align}
Therefore, to transform into the harmonic gauge, one needs to solve a wave equation for each $\xi^0$ and $\xi$. Such solutions do exist for the bouncing background, hence the harmonic gauge is well defined throughout the bouncing phase. Note also that the harmonic gauge has a residual gauge freedom allowed by homogeneous solutions to the wave equations for $\xi^0$ and $\xi$.

The Einstein tensor $G^\mu_{\ \nu}$ is given by, up to first order in perturbations,
\begin{align}
G^0_{\ 0} &= - \frac{3}{a^6} \mathcal{H}^2 + \frac{2}{a^6} \Big[ 3 \mathcal{H} (\psi' + \mathcal{H} A) - \nabla^2 (a^4 \psi + \mathcal{H} \sigma) \Big] \, , \\[4pt]
G^0_{\ i} &= \frac{2}{a^6} \Big[ - \psi' - \mathcal{H} A \Big]_{,i} \, , \\[4pt]
G^i_{\;j} &= \frac{1}{a^6} (-2 \mathcal{H}' + 3 \mathcal{H}^2) + \frac{1}{a^6} \Big[ a^4 \psi - a^4 A + \sigma' \Big]^{,i}_{\ ,j} \nonumber \\*
&\quad + \frac{2}{a^6} \Big[ (\psi' + \mathcal{H} A)' + (\mathcal{H}' - 3 \mathcal{H}^2) A - \frac{1}{2} \nabla^2 \big( a^4 \psi - a^4 A + \sigma' \big) \Big] \delta^i_{\;j} \, .
\end{align}
Here $\sigma$ is the shear perturbation, $\sigma \equiv E' - a^2 B$. Then for the scalar fields with perturbations $\delta\phi$ and $\delta\chi$, the stress energy tensor given by the Lagrangian (\ref{eq:lagrangian}) is, up to first order,
\begin{align}
T^0_{\ 0} &= - \frac{1}{a^6} \Big[ \tfrac{1}{2} ({\phi'}^2 - {\chi'}^2) + a^6 V \Big] - \frac{1}{a^6} \Big[ - ({\phi'}^2 - {\chi'}^2) A + ({\phi'} \delta\phi' - {\chi'} \delta\chi') + a^6 V_{,\phi} \delta\phi \Big] \, , \\[4pt]
T^0_{\ i} &= - \frac{1}{a^6} \big( \phi' \delta\phi - \chi' \delta\chi \big)_{,i} \, , \\[4pt]
T^i_{\ j} &= \frac{1}{a^6} \Big[ \tfrac{1}{2} ({\phi'}^2 - {\chi'}^2) - a^6 V \Big] \delta^i_{\;j} + \frac{1}{a^6} \Big[ ({\phi'} \delta\phi' - {\chi'} \delta\chi') - ({\phi'}^2 - {\chi'}^2) A - a^6 V_{,\phi} \delta\phi \Big] \delta^i_{\;j} \, .
\end{align}
Setting $G^\mu_{\ \nu} = T^\mu_{\ \nu}$ gives the linearized equations
\begin{align}
& 3 \mathcal{H} \psi' + \mathcal{H}' A - a^4 \nabla^2 \psi - \mathcal{H} \nabla^2 \sigma = - \tfrac{1}{2}(\phi' \delta\phi' - \chi' \delta\chi') - \tfrac{1}{2} a^6 V_{,\phi} \delta\phi \, , \label{eq:e1} \\[4pt]
& \psi' + \mathcal{H} A = \tfrac{1}{2}(\phi' \delta\phi - \chi' \delta\chi) \, , \label{eq:e2} \\[4pt]
& \sigma' + a^4 \psi - a^4 A = 0 \, , \label{eq:e3} \\[4pt]
& \psi'' + \mathcal{H} A' + \mathcal{H}' A = \tfrac{1}{2}(\phi' \delta\phi' - \chi' \delta\chi') - \tfrac{1}{2} a^6 V_{,\phi} \delta\phi \, . \label{eq:e4}
\end{align}
In addition, the equations of motion for $\delta\phi$ and $\delta\chi$ are
\begin{align}
& \delta\phi'' - a^4 \nabla^2 \delta\phi + a^6 V_{,\phi\phi} \delta\phi + 2 a^6 V_{,\phi} A - \phi' ( A' + 3 \psi' - \nabla^2 \sigma ) = 0 \, , \\[4pt]
& \delta\chi'' - a^4 \nabla^2 \delta\chi - \chi' ( A' + 3 \psi' - \nabla^2 \sigma ) = 0 \, .
\end{align}
Eq.~(\ref{eq:e4}) is redundant since it can be derived from (\ref{eq:e2}). Eq.~(\ref{eq:e1}) serves as the Hamiltonian constraint, whereas Eq.~(\ref{eq:e2}) is the momentum constraint.

Specifying to the harmonic gauge, Eq.~(\ref{eq:linearharm}) becomes a dynamical equation for $A$, and Eq.~(\ref{eq:linearharmspatial}) for $B$. The complete set of equations are then given by, for a single Fourier mode with wavenumber $k$,
\begin{align}
& A' + 3 \psi' + k^2 (E' - a^2 B) = 0 \, , \label{eq:lineomA} \\[4pt]
& B' + 2 \mathcal{H} B + a^2 (A - \psi + k^2 E) = 0 \, , \label{eq:lineomB} \\[4pt]
& \psi' + \mathcal{H} A = \tfrac{1}{2} (\phi' \delta\phi - \chi' \delta\chi) \, , \label{eq:lineompsi} \\[4pt]
& E'' + a^4 k^2 E = 0 \, , \label{eq:lineomE} \\[4pt]
& \delta\phi'' + a^4 k^2 \delta\phi + a^6 V_{,\phi\phi} \delta\phi + 2 a^6 V_{,\phi} A = 0 \, , \label{eq:lineomphi} \\[4pt]
& \delta\chi'' + a^4 k^2 \delta\chi = 0 \, , \label{eq:lineomchi}
\end{align}
with a constraint coming from Eqs.~(\ref{eq:e1}, \ref{eq:e2}),
\begin{equation} \label{eq:linconstraint}
-\tfrac{1}{2} (\phi'^2 - \chi'^2) A + a^4 k^2 \psi + \mathcal{H} k^2 (E' - a^2 B) + \tfrac{1}{2}(\phi' \delta\phi' - \chi' \delta\chi') + \tfrac{1}{2} a^6 V_{,\phi} \delta\phi + \tfrac{3}{2} \mathcal{H} (\phi' \delta\phi - \chi' \delta\chi) = 0 .
\end{equation}

The initial values for $\delta\phi$, $\delta\phi'$, $\delta\chi$, $\delta\chi'$, $A$, $B$, $\psi$, $E$, and $E'$ are chosen to agree with the initial data for our numerical computations, Eqs.~(\ref{eq:init_g00} - \ref{eq:init_Pchi}). At linear order,
\begin{alignat}{2}
A(0) &= 0, \quad & B(0) &= 0 , \label{eq:linearAmp1} \\[4pt]
\psi(0) &= - 2 \delta\Psi , \quad & E(0) &= 0 , \label{eq:linearAmp2} \\[4pt]
\delta\phi(0) &= f_1 , \quad & \delta\phi'(0) &= a_0^3 ( f_0 - 6 \dot{\phi}_0 \delta\Psi ) , \label{eq:linearAmp3} \\[4pt]
\delta\chi(0) &= f_3 , \quad & \delta\chi'(0) &= a_0^3 ( f_2 - 6 \dot{\chi}_0 \delta\Psi ) , \label{eq:linearAmp4}
\end{alignat}
and $E'(0)$ is given by the constraint equation (\ref{eq:linconstraint}). Here $\delta\Psi$ is given by the conformal factor $\Psi$ in Eq.~(\ref{eq:hamconstr}) expanded to linear order,
\begin{equation} \label{eq:linearPsi}
\delta\Psi = \frac{\dot{\phi}_0 f_0 - \dot{\chi}_0 f_2 - c V_0 f_1}{4 (9 H_0^2 - 3 V_0 + k^2 / a_0^2)} \, ,
\end{equation}
and the parameters $f_0$ - $f_3$ are specified according to our numerical computations. Note that the above initial values satisfy the relation
\begin{equation} \label{eq:CMClin}
\psi(0) = - \frac{1}{2 a_0 k^2} \big( \dot{\phi}_0 \delta\phi'(0) - \dot{\chi}_0 \delta\chi'(0) + a_0^3 V'_0 \delta\phi(0) \big) \, ,
\end{equation}
which agrees with the constant mean curvature initial data in our numerical computations.

\section{Bunch-Davies initial values} \label{sec:initial}

The adiabatic perturbations arise from quantum fluctuations when the modes are deep inside the horizon. The scalar field perturbations can be studied by using the canonical variables $u_\phi = a \, \delta\phi$ and $u_\chi = a \, \delta\chi$ \cite{Mukhanov:1990me}, which satisfy the equation
\begin{equation} \label{eq:mukhanov}
u'' - \frac{a''}{a} \, u + k^2 u = 0 ,
\end{equation}
where in this equation $'$ denotes the derivative with respect to the conformal time $\eta$, defined by $d\eta = d\tau / a$. For a matter-like contraction phase with $w = 0$, the scale factor $a$ follows a power law solution $a \sim (-\tau)^{2/3} \sim (-\eta)^2$, hence $a''/a = 2/\eta^2$. When the mode is deep inside the horizon, $k (-\eta) \gg 1$, the quantum fluctuations should match the Bunch-Davies vacuum state,
\begin{equation}
u \to \frac{1}{\sqrt{2k}} \, e^{-i k \eta} \, .
\end{equation}
For this initial condition, the solution to Eq.~(\ref{eq:mukhanov}) is given by
\begin{equation} \label{eq:u=}
u = \sqrt{\frac{\pi x}{4 k}} \, H^{(1)}_\nu (x) \, e^{i (\frac{\nu}{2} + \frac{\pi}{4})} \, ,
\end{equation}
where $x = k (-\eta)$, and $H^{(1)}_\nu$ is the Hankel function of the order $\nu = \frac{3}{2}$. As $x \to 0$ in the late contraction phase, the Hankel function takes the asymptotic form
\begin{equation} \label{eq:Hankel}
H^{(1)}_\nu \to \frac{1}{\Gamma(\nu+1)} \Big( \frac{x}{2} \Big)^\nu - i \, \frac{\Gamma(\nu)}{\pi} \Big( \frac{x}{2} \Big)^{-\nu} \, .
\end{equation}
Neglecting the unsubstantial phase factor in (\ref{eq:u=}), the scalar field perturbation can be written as
\begin{equation} \label{eq:deltaphiflat}
\delta\phi_\psi \sim \frac{1}{3\sqrt{2}} k^{3/2} - \frac{i}{8\sqrt{2}} k^{-3/2} (-\eta)^{-3} \, ,
\end{equation}
and the same for $\delta\chi_\psi$.

Switching to the harmonic time $t$, using the relation $dt = d\eta/a^2$ and hence $(-\eta) \sim (t - t_{-\infty})^{-1/3}$, we can write
\begin{equation} \label{eq:C1+C2}
\delta\phi_\psi = -i C_1(k) (t - t_{-\infty}) + C_2(k) ,
\end{equation}
where $t_{-\infty}$ corresponds to the time when $\eta \to -\infty$. The constants $C_1$, $C_2$ scale as
\begin{equation}
C_1(k) \sim k^{-3/2} , \quad C_2(k) \sim k^{3/2} ,
\end{equation}
and their relative size is fixed by Eq.~(\ref{eq:deltaphiflat}),
\begin{equation} \label{eq:C2overC1}
\frac{C_2}{C_1} = \frac{8}{3} (- k \eta)^3 (\tau - \tau_{-\infty}) = \frac{8}{3} \Big| \frac{k}{aH} \Big|^3 \Big( \frac{-2}{3\mathcal{H}} \Big) \, .
\end{equation}
Therefore, for long wavelengths with $k \ll a H$, the $C_1$ term in Eq.~(\ref{eq:C1+C2}) dominates and gives rise to a scale invariant power spectrum. The size of $C_1$ can be normalized according to the comoving curvature perturbation $\mathcal{R}$ as follows.

The curvature perturbation $\mathcal{R}_\phi$ in the comoving $\phi$ gauge is defined as
\begin{equation} \label{eq:Rphi}
\mathcal{R}_\phi \equiv \psi + \mathcal{H} \frac{\delta\phi}{\phi'} \, ,
\end{equation}
which is related to the scalar field perturbation $\delta\phi_\psi$ in the flat gauge by $\mathcal{R}_\phi = \frac{\mathcal{H}}{\phi'} \delta\phi_\psi$. In our computation we consider a finite volume of size $L$ with periodic boundary condition. So the amplitude can be expanded as a Fourier series
\begin{equation}
\mathcal{R}(\mathbf{x}) = \sum_{\mathbf{k}} \mathcal{R}_\mathbf{k} \, e^{i \mathbf{k} \cdot \mathbf{x}} \, ,
\end{equation}
where $k_i = \frac{2\pi}{L} n_i$, $n_i = 0, \pm 1, \cdots$, and
\begin{equation}
\mathcal{R}_\mathbf{k} = \frac{1}{L^3} \int d^3\mathbf{x} \, \mathcal{R}(\mathbf{x}) \, e^{-i \mathbf{k} \cdot \mathbf{x}} \, .
\end{equation}
We can write $\mathcal{R}_\mathbf{k} \equiv a_\mathbf{k} - i b_\mathbf{k}$, where
\begin{align}
a_\mathbf{k} &= \frac{1}{L^3} \int d^3\mathbf{x} \, \mathcal{R}(\mathbf{x}) \cos (\mathbf{k} \cdot \mathbf{x}) \, , \\[4pt]
b_\mathbf{k} &= \frac{1}{L^3} \int d^3\mathbf{x} \, \mathcal{R}(\mathbf{x}) \sin (\mathbf{k} \cdot \mathbf{x}) \, .
\end{align}
The reality of $\mathcal{R}(\mathbf{x})$ implies $a_{-\mathbf{k}} = a_\mathbf{k}$ and $b_{-\mathbf{k}} = - b_\mathbf{k}$. These two coefficients $a_\mathbf{k}$, $b_\mathbf{k}$ nicely correspond to the $C_2$ and $C_1$ terms above. In terms of these coefficients, the real space amplitude can be written as
\begin{equation} \label{eq:cos+sin}
\mathcal{R}(\mathbf{x}) = \sum_\mathbf{k} a_\mathbf{k} \cos (\mathbf{k} \cdot \mathbf{x}) + b_\mathbf{k} \sin (\mathbf{k} \cdot \mathbf{x}) \, .
\end{equation}

In the limit $L \to \infty$, the Fourier transform becomes
\begin{equation}
\mathcal{R}(\mathbf{x}) \to \frac{L^3}{(2\pi)^3} \int d^3\mathbf{k} \, \mathcal{R}_\mathbf{k} \, e^{i \mathbf{k} \cdot \mathbf{x}} \, .
\end{equation}
And the autocorrelation function is given by
\begin{align}
\langle \mathcal{R}(\mathbf{x})^2 \rangle &= \frac{1}{L^3} \int d^3\mathbf{x} \, \mathcal{R}(\mathbf{x}) \mathcal{R}(\mathbf{x})^* = \frac{L^3}{(2\pi)^3} \int d^3\mathbf{k} \, \big| \mathcal{R}_\mathbf{k} \big|^2 = \int d\log k \, \frac{(k L)^3}{2 \pi^2} \big| \mathcal{R}_\mathbf{k} \big|^2 \, .
\end{align}
The power spectrum can be defined as
\begin{equation} \label{eq:power-spectrum}
\Delta_\mathcal{R}^2 \equiv \frac{(k L)^3}{2 \pi^2} \big| \mathcal{R}_\mathbf{k} \big|^2 \, ,
\end{equation}
which has to match the observed nearly scale invariant amplitude $\Delta_\mathcal{R}^2 \approx 2.4 \times 10^{-9}$ \cite{Hinshaw:2012aka}.

\end{document}